\documentclass{aa}  

\usepackage{graphicx}
\usepackage{natbib}
\bibpunct{(}{)}{;}{a}{}{,} 
\usepackage{hyperref}
\hypersetup{colorlinks=true, citecolor=blue}
\usepackage{color}
\usepackage{lscape}
\usepackage{longtable}
\usepackage{times}
\usepackage{nccmath}
\usepackage{float}
\usepackage{xcolor}

\begin{document}

   \title{\textcolor{blue}{ Cold dust and stellar emissions in dust-rich galaxies observed with ALMA: a challenge for SED-fitting techniques}}
   \author{V. Buat
          \inst{1}
          \and L. Ciesla
          \inst{1} 
          \and M. Boquien \inst{2} 
         \and  Ma\l{}ek, K. \inst{1,3}
          \and D. Burgarella \inst{1}
          }

    \institute{Aix Marseille Univ, CNRS, CNES, LAM Marseille, France
              \email{ veronique.buat@lam.fr}
                 \and Centro de Astronom\'ia (CITEVA), Universidad de Antofagasta, Avenida Angamos 601, Antofagasta, Chile 
                   \and National Centre for Nuclear Research, ul. Hoza 69, 00-681 Warszawa, Poland
             }



  \abstract
   {Over the past few years the Atacama Large Millimeter Array (ALMA) has detected dust-rich galaxies whose cold dust emission is spatially disconnected from the ultraviolet (UV) rest-frame emission. This represents a challenge for modeling their spectral energy distributions (SED) with codes based on an energy budget between the stellar and dust components. This could potentially weaken the solidity of the physical parameters measured with these modeling tools.}
   {We want to verify  the validity of energy balance modeling on a sample of galaxies observed from the UV to the sub-millimeter rest frame with ALMA and decipher what information can be reliably retrieved from the analysis of the full SED and from subsets of wavelengths.}
   {We select 17 sources at $z\simeq 2$ in the {\sl Hubble} Ultra-Deep Field (HUDF) and in the GOODS-{\sl South} field   detected with ALMA and {\sl Herschel} and for which UV to near-infrared rest-frame ancillary data are available. We fit the data with CIGALE exploring different configurations for dust attenuation and star formation histories, considering either the full dataset or one that is reduced to the stellar and dust emission. We compare estimates of the dust luminosities, star formation rates, and stellar masses.}
  {The fit of the stellar continuum alone with the starburst attenuation law can only reproduce up to 50\% of the total dust luminosity observed by {\sl Herschel} and ALMA. This deficit is found to be marginally consistent with  similar quantities estimated in the COSMOS field and is found to increase with the specific star formation rate. The combined stellar and dust SEDs are well fitted when different attenuation laws are introduced. Shallow attenuation curves are needed for the galaxies whose cold dust distribution is very compact compared to starlight.  The stellar mass estimates are affected by the choice of the attenuation law. The star formation rates are robustly estimated as long as dust luminosities are available. The large majority of the galaxies are above the average main sequence of star forming galaxies and one source is a strong starburst.}
   {}

   \titlerunning{ SED fitting of ALMA galaxies}
   \maketitle
 %

\section{Introduction}

Modeling the  spectral energy distributions (SED) of galaxies is commonly used to measure some of the most important physical parameters that quantify galaxy evolution such as the star formation rate (SFR) and the stellar mass. The central idea behind this method is to reconstruct the stellar emission of a galaxy with population-synthesis models, assuming star formation histories of varying complexity \citep[e.g.,][]{Walcher11, Conroy13}. Dust however plays a crucial role by strongly affecting and reshaping the SED: it absorbs and scatters stellar photons and thermally emits the absorbed energy in the infrared (IR; $\lambda \sim 1-1000~\mu$m). To account for the impact of dust, modern physical models link the stellar and dust components: the energy radiated by dust corresponds to the absorbed stellar light.

This energy budget is implicit in radiation transfer models aimed at modeling nearby, spatially resolved galaxies with more or less complex geometries \citep[e.g.,][]{DeLooze14}. It is  implemented explicitly in a handful of panchromatic SED fitting codes aimed at fitting the SED of galaxies such as MAGPHYS \citep{daCunha08}, CIGALE \citep{Noll09, Boquien19}, PROSPECTOR \citep{Leja17}, and BAGPIPES \citep{Carnall18}. These SED fitting codes are all based on simplified assumptions; for example, the emission is supposed to be isotropic at all wavelengths, and the effect of dust on the stellar continuum (and nebular emission for BAGPIPES and CIGALE) is accounted for by applying an effective attenuation law. This curve is either measured as for the commonly used starburst law \citep{Calzetti00} or built to reproduce average observed trends \citep{Charlot00, LoFaro17, Cullen18}. These methods are very efficient at fitting very large samples of galaxies. 
\citet{Hayward15} showed that  the SED modeling code MAGPHYS recovers physical parameters reasonably well for most of their simulated galaxies.

It has recently been shown that effective attenuation laws are far from universal but actually vary in local and more distant galaxies \citep[e.g.,][]{Salmon16, Salim18, Buat18}. The attenuation law appears to flatten out when the amount of obscuration increases, a trend that is predicted by radiation transfer models in dusty media  with idealized geometries \citep[][and references therein]{Chevallard13} or more realistic distributions from hydrodynamical simulations \citep[][]{Jonsson10, Roebuck19, Trayford19}.  \citet{LoFaro17} found that a very flat attenuation law accurately describes a sample of ultra luminous infrared galaxies  (ULIRGs) at $z\sim 2$. Such a curve is able to explain the locus of these galaxies in the IRX-$\beta$ diagram\footnote{This diagram relates the slope $\beta$ of the UV continuum to IRX defined as the ratio of the IR and UV luminosities.}, well above that of starburst galaxies obtained by \citet{Meurer99}. In this case high IRX values for a given $\beta$ are reproduced by a global attenuation law that is grayer than the starburst curve rather than being due to a disconnection between fully obscured star-forming components and unobscured regions dominating the UV emission.

Energy balance methods assume a physical coupling between the dust and stellar components and should in principle only be valid in this case. Infrared-bright galaxies in which most of the dust emission is not coincident with the emergent stellar emission have long since been identified however \citep[e.g.,][]{Charmandaris04, Howell10}. More recently,  observations of high-redshift massive dusty and submillimeter (submm) galaxies with the Atacama Large Millimeter Array (ALMA) report a very compact cold dust emission, with a rest-frame UV emission distributed in clumps that are spatially distant from the compact dust emission, and an extended H-band distribution \citep[e.g.,][]{Hodge16, Tadaki17, GomezGuijarro18, Puglisi19}. These dusty, compact objects could represent $\sim 50\%$ of the massive ($M > 10^{11} M_\sun$) main sequence (MS) population \citep[][]{Puglisi19}. Conversely, by stacking ALMA images of less massive star-forming galaxies at $z\sim 2$, \citet{Lindroos16} found typical sizes of $\sim 5 $ kpc for the ALMA continuum, similar to the median sizes measured in the near-infrared (NIR). Using observations  of galaxies with vigorous star formation at $z= 1-3$ with both the Kark G. Jansky Very Large Array and ALMA, \citet{Rujopakarn16} also report star formation distributed over an area similar to the distribution of stellar mass, although ALMA high-resolution maps  picture  complex configurations for a few MS galaxies \citep{Rujopakarn19}.
 
Very different dust and stellar distributions obviously challenge a local energy balance which is nevertheless expected to be valid at a global scale provided that we are dealing with a single physical object where dust emission comes from heating by stellar photons. However the way in which both emissions are linked,  assuming a  simple star formation history (SFH)  and  dust attenuation recipe, can have an impact on the physical parameters derived from the modeling of the global emission. In this paper, we analyze a sample of $z \sim 2$ galaxies observed with ALMA and complemented by a rich multi-wavelength dataset from the rest-frame UV to the  submm. We explore which SFHs and attenuation laws are needed to fit the SED of these galaxies and to what extent the SFR and stellar masses determined from the fit, with and without the IR emission emitted by dust, are reliable. With this aim in mind, we use data from the {\sl Hubble} Ultra-Deep Field (HUDF) \citep{Dunlop17}  GOODS-{\sl South} field  \citep{Elbaz18} surveys. We use CIGALE \citep{Boquien19} to fit the SED of these objects with three different attenuation laws that are representative of the very different shapes of effective attenuation laws used in the literature: the starburst law of \citet{Calzetti00}, and the recipes of \citet{Charlot00} and \citet{LoFaro17}.

The paper is organized as follows. The galaxy sample and the multi-wavelength data are presented in section \ref{sec:data}. The different configurations of the SED fitting process are explained in section \ref{sec:method} and the fits are performed and discussed in section \ref{sec:fit}. In sections \ref{sec:dust} and \ref{sec:sfr-mstar} we discuss the measures of the dust luminosity ($\rm L_{dust}$), SFR, and stellar mass, and the locus of the galaxies on the MS. The dust and stellar relative distributions are related to the attenuation law in section \ref{sec:laws}. We conclude in section \ref{sec:conclusion}.

\section{Target selection and multi-wavelength data\label{sec:data}}
We consider the targets from the samples of \citet{Elbaz18} and \citet{Dunlop17}, hereafter respectively labeled GS and UDF. We first select galaxies with ALMA observations corresponding to rest-frame wavelengths $\le 500 ~\mu$m. This corresponds to targets with a redshift $z>1.6$ for the sources of \citet{Dunlop17} observed at 1.3~mm and to $z>0.74$ for the sample of \citet{Elbaz18} observed at 870~$\mu$m. We are left with ten objects from \citet{Dunlop17} and seven sources from \citet{Elbaz18} (we do not include UDF12 listed with a redshift $z=5$ and the optically dark galaxy GS8 at $z=3.2$). All the sources but two (UDF10 and UDF15) are part of the extended sample analyzed by \citet{Elbaz18} and ten sources are detected in X-ray. The redshifts range from 1.6 to 2.8 with a median value at 2.3. The  identifiers, redshifts of the sources (with the references), and X-ray detections if any are listed in Table \ref{attlaws}.  

The objects have been observed from the rest-frame UV to the far-IR. From the shortest to the longest wavelengths: HST/ACS (F435W, F606W, F775W, F814W, F850LP) and HST/WFC3 (F105W, F125W, F160W), the VLT/ISAAC Ks band,  {\it Spitzer}/IRAC (four bands), the \textit{Spitzer}/MIPS band at 24$\rm \mu m$, and \textit{Herschel}/PACS and SPIRE (five bands). 
The optical to IRAC data come from the multi-wavelength catalog of the Cosmic Assembly Near-infrared Deep Extragalactic Legacy Survey  \citep[CANDELS,][]{Guo13}. We cross-match the coordinates of the ALMA sources with those of the CANDELS catalog with a tolerance radius of 1 arcsec, confirming the associations performed by \citet{Elbaz18}, except for UDF 10 and 15 which are not included in the \citet{Elbaz18} sample. The data from the {\it Herschel} GOODS-{\sl South} field catalog were produced by the ASTRODEEP team\footnote{\url{http://www.astrodeep.eu/category/data/data-release/}} (Wang et al., in preparation). Postage-stamp HST  images with ALMA contours for each source are shown in \citet{Dunlop17}, \citet{Rujopakarn16} and \citet{Elbaz18}.
 
\section{The SED fitting method\label{sec:method}}
\subsection{The code CIGALE}

The SED fitting is performed with CIGALE\footnote{\url{https://cigale.lam.fr}}. We refer to \citet{Boquien19} for a detailed description of the code. CIGALE combines a stellar SED with dust attenuation and emission components. A critical aspect is that it conserves the energy between dust absorption in the UV-to-NIR domain and emission in the mid-IR and far-IR. The quality of the fit is assessed by the best $\chi^2$ (and a reduced best $\chi^2$ defined as $\rm \chi_r^2= \chi^2/\left(N-1\right)$, with N being the number of data points). The value of the physical properties and their corresponding uncertainties are estimated as the likelihood-weighted means and standard deviations. In this work, we introduce different attenuation models. To compare them we use the Bayesian inference criterion defined as $\mathrm{BIC}= \chi^2+k \times \ln(N)$, with $k$ being the number of free parameters and $N$ the number of data points used for the fit \citep{Ciesla18}. The BIC is an approximation of the Bayes factor \citep{Kass95}.
Below, we describe the assumptions and choices adopted to run CIGALE for our current study. The values of the input parameters are listed in Table \ref{param}.

\subsection{Star formation histories, dust, and nebular components}
We assume a delayed SFH with the functional form $\textrm{SFR} \propto t \times \exp(-t/\tau$). A current burst of star formation is introduced if it provides better fits. This burst is modeled as a constant star formation that has been ongoing for at most 100~Myr and that is superimposed to the delayed SFH. Its amplitude is measured by the stellar mass fraction produced during the burst episode. The age of the main stellar population and of the burst are also free parameters as well as the e-folding timescale $\tau$; the input values are listed in Table \ref{param}. We adopt a \citet{Chabrier03} initial mass function (IMF) and the stellar models of \citet{bc03}. The metallicity is fixed to the solar value ($\textrm{Z}=0.02$). The input values characterizing the SFH are presented in Table \ref{param}. Here, $\rm L_{dust}$ is measured with the energy budget and the IR SED corresponding to the dust component is calculated with the \citet{Draine07} models with the same input parameters as in \citet{LoFaro17}. For the sources  detected in X-ray an AGN component is added to the fit following the method developed in \citet{Buat15}: the \citet{Fritz06} library of CIGALE is reduced to two AGN models with an angle between the AGN axis and the line of sight equal to 80$^{\circ}$ and either a small or high optical depth at 9.7 $\mu$m. The nebular emission is added  from the  Lyman continuum photons produced by  the stellar component. The nebular continuum and the emission lines are calculated from a grid of nebular templates generated with CLOUDY 08.0. In this work the ionization parameter is chosen to be $\rm log(U)=-2$.
\begin{table*}
\scriptsize
\centering
\begin{tabular}{l c c}
\hline\hline
Parameter & Symbol & Range \\
\hline
\multicolumn{3}{c}{Delayed SFH and recent burst}\\
\hline
age of the main population& $age_{main}$ & 1500, 2000, 3000, 4000\,Myr\\
$e$-folding timescale of the delayed SFH & $\tau$ & 200,500,1000,2000\,Myr\\
age of the burst & $age_{burst}$ & 5, 10, 50, 100\,Myr\\
burst stellar mass fraction & $f_{burst}$& 0.0, 0.1, 0.15,0.2,0.25\\
\hline
\multicolumn{3}{c}{Dust attenuation}\\
\hline
\textit{ C00 law:} & &\\
color excess & $E(B-V)$ & 0.1-1\\
\hline
\textit{CF00 and LF17 laws:} & &\\
V-band attenuation in the ISM & $A_{\rm V}^{\rm ISM}$ & 0.3-5\\
power law slope of dust attenuation in the BCs & $n^{\rm BC}$ & -0.7\\
$A_{\rm V}^{\rm ISM}/ (A_{\rm V}^{\rm ISM}+A_{\rm V}^{\rm BC}$) & $ \mu$ & 0.5 \\
power law slope of dust attenuation in the ISM & $n^{\rm ISM}$ & -0.7 (CF00), -0.48 (LF17)\\
\hline
\multicolumn{3}{c}{Dust emission}\\
\hline
mass fraction of PAH & $q_{\rm PAH}$ & 1.12, 2.50, 3.19\\
minimum radiation field & U$_{\rm min}$ & 5., 10., 25., 40.\\
powerlaw slope dU/dM $\propto$ U$^{\alpha}$ & $\alpha$ & 2.0\\
dust fraction in PDRs & $\gamma$ & 0.02\\
\hline
\multicolumn{3}{c}{AGN emission}\\
\hline
optical depth & $\tau$ & 1,6\\
AGN luminosity fraction & $f_{AGN}$&0,0.1,0.2,0.3\\
\hline\hline
\end{tabular}
\caption{CIGALE modules and input parameters used for all the fits. The IMF is that of \citet{Chabrier03} with $\textrm{Z}=0.02$.}
\label{param}
\end{table*}

\subsection{Attenuation laws}
We expect the diverse and complex structures of our targets to translate to different effective attenuations. Three different dust attenuation recipes  already used in the literature are considered. We do not introduce full flexibility: the large majority of studies using SED fitting methods assume a fixed law. Our aim in this work is to discuss the reliability of physical parameter estimations under  simple classical approaches.

\begin{figure}
\begin{center}
\includegraphics[width=8cm] {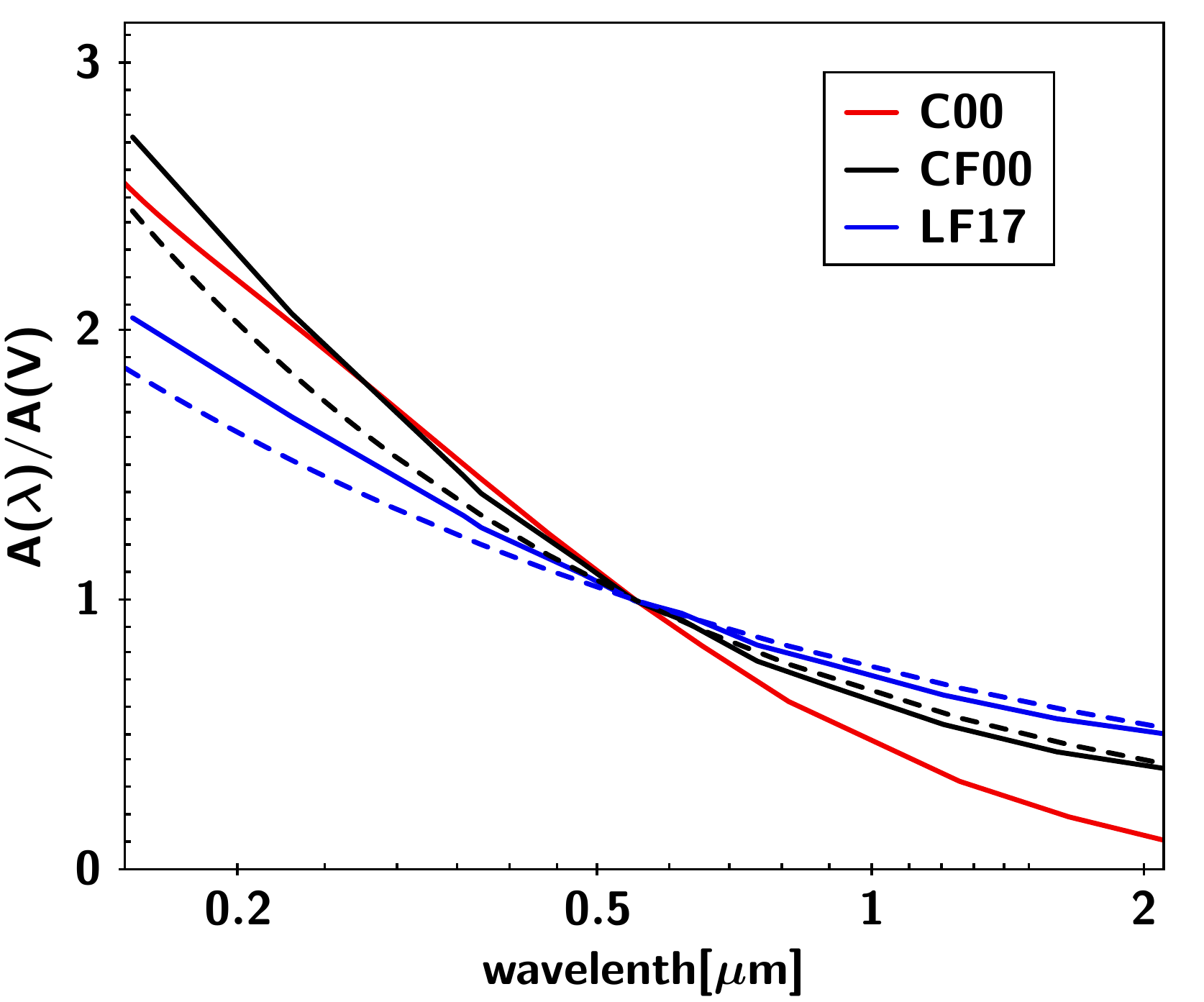}
\end{center}
\caption{Shapes of the attenuation laws introduced for the modeling. The ISM power-laws for CF00 (black) and LF17 (blue) are plotted with dotted lines. The solid lines represent the C00 (red)  law and the average attenuation laws obtained for our sample with CF00 and LF17; see text for details.}
\label{dustlaw}
\end{figure} 

First, we consider the attenuation law measured for the stellar continuum of nearby starbursts by \citet{Calzetti00}, for which the amount of attenuation is quantified with the color excess $A(\lambda) = E({\rm B-V})_{\rm star} k^{\prime}(\lambda)$. The nebular emission is attenuated with a simple screen model and a Milky Way extinction curve with a color excess $E({\rm B-V})_{\rm line}$.  The ratio between the color excess for the stellar continuum and the nebular lines was originally measured to be equal to 0.44 \citep{Calzetti01} in local starbursts. At $z\ge 1,$ \citet{Kashino13, Puglisi16}  obtained  more similar  attenuations for stellar and nebular emissions and \citet{Buat18} concluded that there is great variability among galaxies with an average value of 0.55. Our targets are  highly obscured, with a  compact and dusty star formation. Young stars emitting in the UV are expected to experience very high attenuation and we do not have the measurement of emission lines to constrain the differential attenuation. We first tested three different values of the ratio, namely 0.44, 0.6,  and 1, with no significant change in the results. We  decided to use the same color excess for stellar and nebular emission. The recipe  is  referred to as C00 hereafter.

Here, we introduce the  attenuation recipe of \citet{Charlot00}, which differs from C00 in its philosophy. A differential attenuation between young (age $<10^7$ years) and old (age $>10^7$ years) stars is assumed. Light from both young and old stellar populations is attenuated in the interstellar medium (ISM) but light from young stars is also affected by an extra attenuation in the birth clouds (BCs). Both attenuation laws are modeled by a power law and the amount of attenuation is quantified by the attenuation in the V band, $A^{\rm BC}_{\lambda} = A^{\rm BC}_{\rm V} (\lambda/0.55 )^{n^{\rm BC}}$ and $A^{\rm ISM}_{\lambda} = A^{\rm ISM}_{\rm V} (\lambda/0.55 )^{n^{\rm ISM}}$. \citet{Charlot00} fixed both exponents of the power laws $n^{\rm BC}$ and $n^{\rm ISM}$ to $-0.7$, although a value of $-1.3$ is initially introduced in their model and further adopted by \citet{daCunha08}. Here, we use the original scenario which is  more consistent with the studies of HII regions \citep{LoFaro17}. The $\mu$ parameter is defined as the ratio of the attenuation in the V band experienced by old and young stars $\mu = A^{\rm ISM}_{\rm V} /( A^{\rm ISM}_{\rm V} +A^{\rm BC}_{\rm V})$. \citet{Charlot00} obtained $\mu = 0.3$ from their study of nearby starburst galaxies. From a study of several thousand galaxies detected by Herschel as part of the HELP project, \citet{Malek18} found that a $\mu$ value slightly higher than $0.3$ led to better fits. The recipe defined with the exponents of the power laws $n^{\rm BC}$ and $n^{\rm ISM}$ equal to $-0.7$ and $\mu$ equal to $0.5$ is referred to as CF00 hereafter.

We further introduce the modification to  CF00 as proposed by \citet{LoFaro17} to fit ULIRGs at $z\simeq 2$ in agreement with radiation transfer modeling. This attenuation law is flatter with $n^{\rm ISM}=-0.48$. We refer to this third law as LF17 hereafter. We use the same value of $\mu= 0.5$ in this case, in agreement with the results of \citet{LoFaro17}.

We compare these three laws in Fig.~\ref{dustlaw}. While the shape of C00 is fixed and applies uniformly to the stellar continuum, the effective shape of CF00 and LF17 instead depends on the SFH since the young and old stars experience a different attenuation. In Fig.~\ref{dustlaw} we report the average law for our galaxy sample from the fits performed in section \ref{sec:fit}. We see that CF00 combined with the SFH obtained for our sample leads to an attenuation similar to C00 at short wavelengths. This is expected since \citet{Charlot00} built their model to reproduce the properties of the starburst galaxies analyzed by \citet{Calzetti94}. However a significant difference is visible for $\lambda>0.5~\mu$m: CF00 is much flatter than C00 \citep{Chevallard13,LoFaro17, Buat18}. The average LF17 law is flatter than the two previous laws over the full wavelength range.
 
\section{Fitting the SEDs\label{sec:fit}}
The usual way to model galaxies is to use all the photometric data available over the broadest feasible wavelength range in order to measure the physical parameters as reliably as possible. In the particular case of the galaxies studied in this work, the different locations of dust and stellar emissions lead us to explore different configurations for the fits. We perform three fits here: stellar continuum only, dust emission only, and finally the full SED.

\subsection{Fitting the stellar continuum}
We model the stellar continuum using all the HST, VLT, and IRAC data available.  We run CIGALE with the three attenuation laws and with and without a burst added to the delayed SFH.

First we test the requirement to add a burst to the delayed SFH. The comparison of the BIC is used for this purpose. Two additional free parameters are introduced when a burst is added to the delayed SFH. The difference between the two BICs corresponding to the models with and without a burst is calculated as $\rm \Delta BIC= \chi^2(burst)+2 \times \ln(N) -\chi^2(no~ burst)$, with N being the number of fitted bands. We adopt the limit of $\rm |\Delta BIC|>6$ to put a preference on a model: this represents strong evidence against the model with the higher BIC \citep{Kass95, Salmon16}. In all cases, $\rm\Delta BIC$ is found between $+5$ and $-3$. There is no strong evidence for a recent burst and a simple delayed SFH is adopted for all the fits of the stellar continuum.

\begin{figure}
\includegraphics[width=8cm]{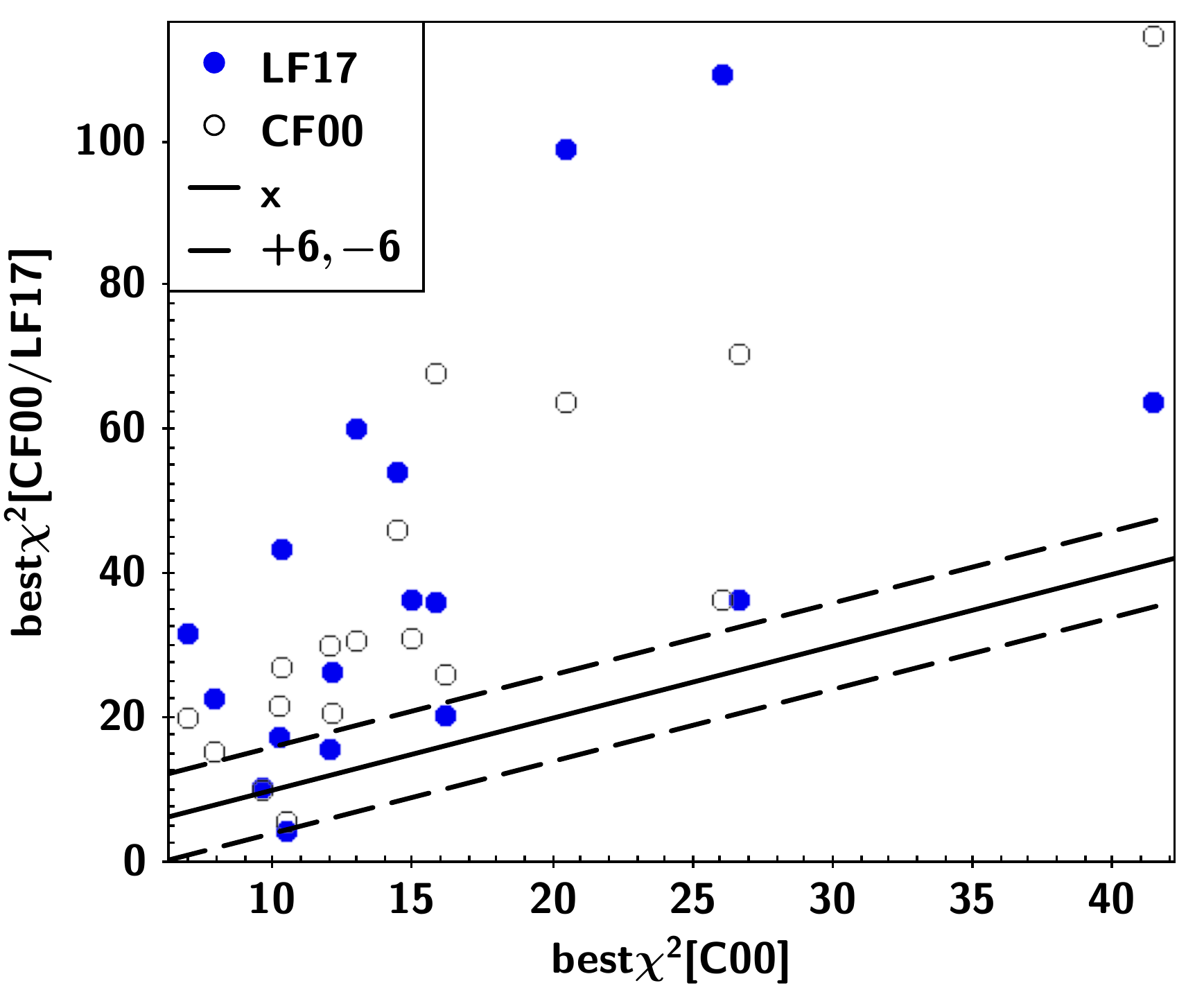}
\caption{Comparison between  best $\chi^2$ obtained with each of the three attenuation laws when only the  stellar continuum is fitted. The values for C00 are plotted on the x-axis against the values for CF00 (empty circles) and LF17 (blue filled circles) on the y-axis. The solid line represents the 1:1 relation and the dashed lines correspond to $\rm |\Delta BIC|=6$.}
\label{chi2-noir}
\end{figure}

We then compare the three attenuation scenarios. They have the same number of free parameters and $\rm \Delta BIC$ is now defined as the difference between the best $\chi^2$ of each law: $\rm \Delta BIC = \chi^2(C00)-\chi^2 (CF00~ or ~ LF17)$. The $\chi^2$ are plotted in Fig.~\ref{chi2-noir} with limits corresponding to $\rm |\Delta BIC|>6$ . We clearly see that C00 gives the best fits. In all cases, C00 is acceptable with $\rm \Delta BIC<6$ ($\rm \Delta BIC$ is even lower or equal to zero except in one case). We adopt the C00 dust attenuation scenario for all our galaxies. The best fits of the stellar continuum for the whole galaxy sample are shown in Appendix A.
 
\subsection{Fitting the dust IR SED}
\begin{figure}
\includegraphics[width=8cm] {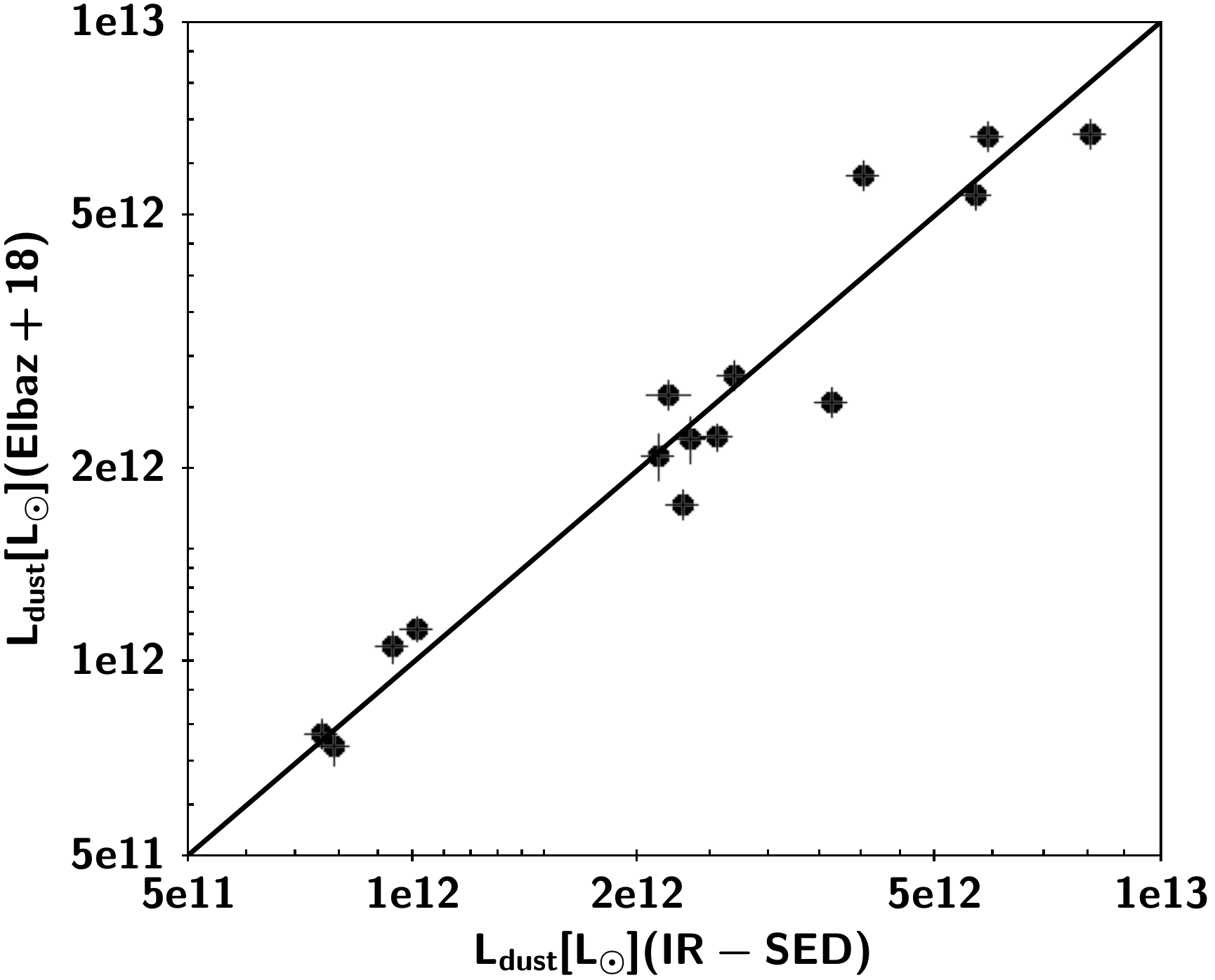}
\caption{Comparison between  $\rm L_{dust}$ measured from the fit of the MIPS-to-ALMA SED with CIGALE (x-axis) and the measurements of  \citet{Elbaz18} (y-axis).}
\label{Ldust-IR}
\end{figure}

We model the dust emission using all the IR data available ({\it Spitzer}/MIPS, {\it Herschel}/PACS and SPIRE, and ALMA) to measure $\rm L_{dust}$ independently of the stellar continuum. This quantity is commonly used to measure obscured star formation using simple assumptions about the SFH and dust absorption \citep[e.g.,][]{Kennicutt12}.  In this case no energy budget is performed and the IR data are fitted to the  \citet{Draine07} models.  No AGN is introduced. Without IRAC data which  are not considered for the fits, the constraint on the AGN component would be very weak. We plot the resulting values in Fig.~\ref{Ldust-IR} and compare them to the measurements of \citet{Elbaz18} for the 15 galaxies in common. We find very good agreement between both values. The $\rm L_{dust}$ values are listed in Table \ref{bayesparam}.

\subsection{Fitting the full SED}
\begin{figure}
\includegraphics[width=8cm] {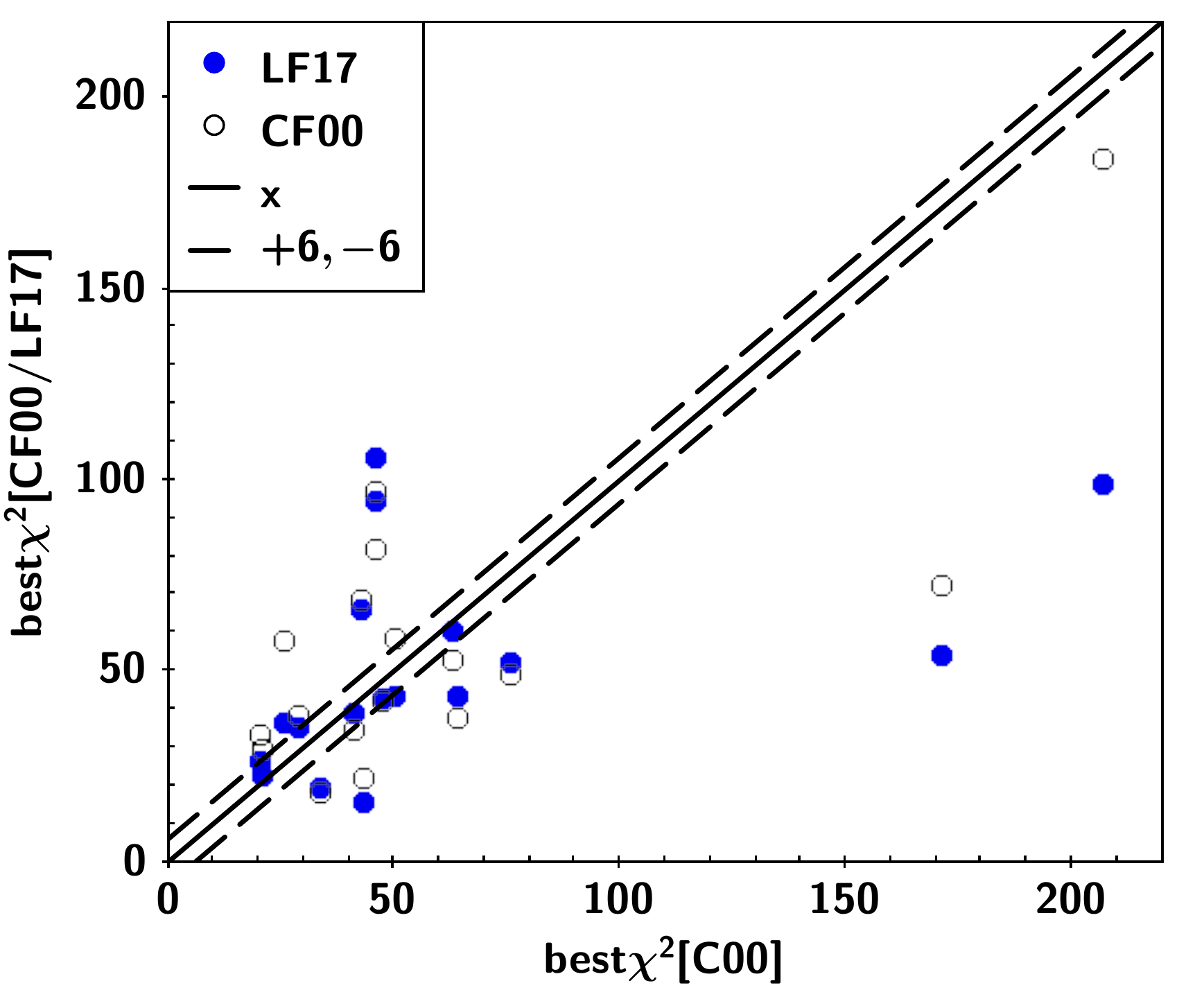}
\caption{Comparison between  best $\chi^2$ obtained with each of the three attenuation laws when  the  full SED  is fitted.  Same quantities and symbols as in Fig.\ref{chi2-noir}.}
\label{chi2-ir}
\end{figure}

\begin{figure*}
\includegraphics[width=2\columnwidth]{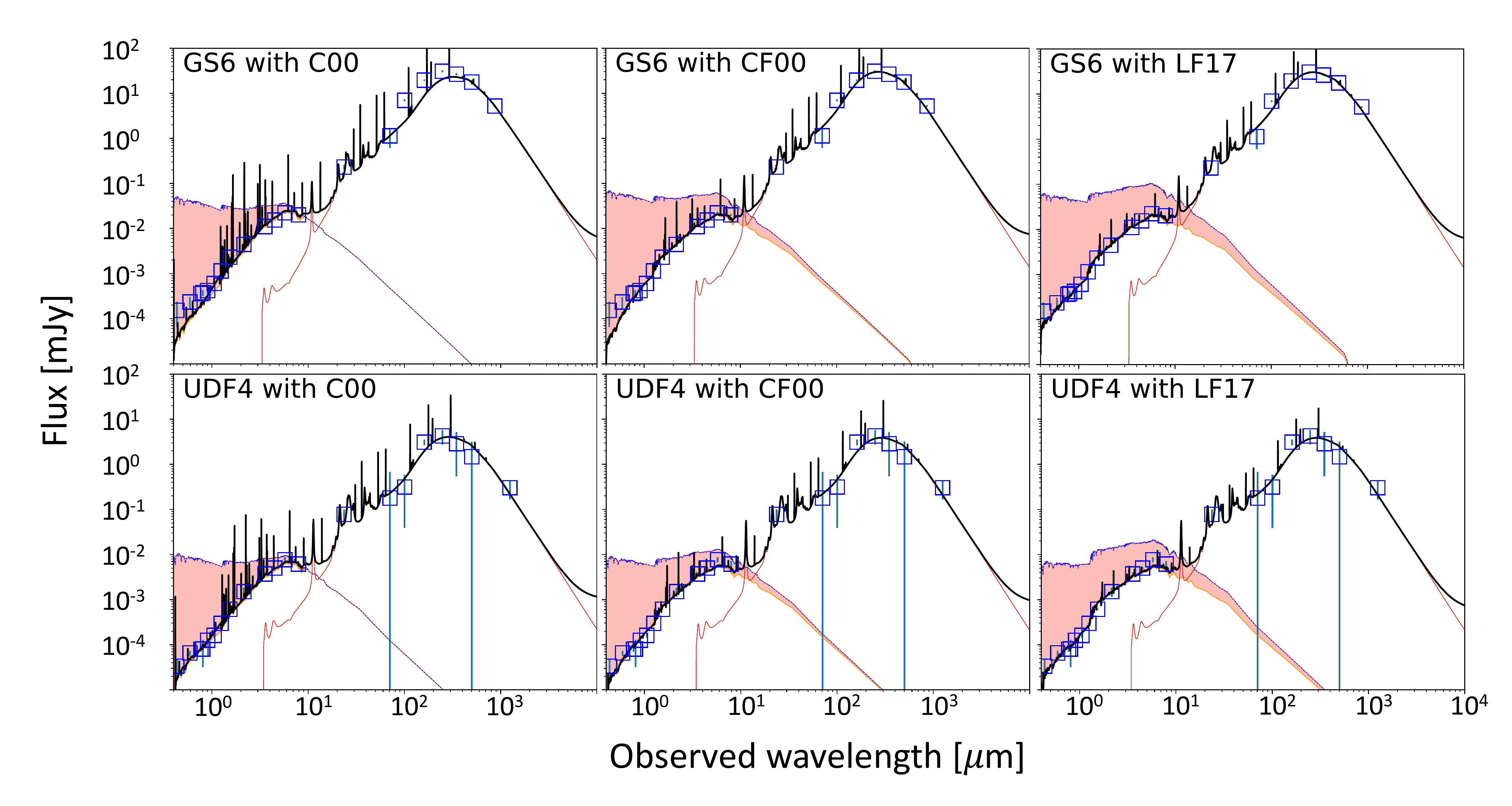}
\caption{Examples of best fits of the full SED  with the three attenuation laws for two galaxies :UDF4 (lower row) and GS6 (upper row). x-axis: observed wavelength in $\mu$m, y-axis: observed flux in mJy. From left to right: C00, CF00, and LF17 laws.  The black line represents  the best fitted spectrum  composed of attenuated stellar emission, nebular lines, and dust emission. The data points are plotted with empty blue squares and their 3 $\sigma$ error with blue vertical lines. The intrinsic, unabsorbed stellar continuum is plotted with a dashed blue line, the red shaded area indicates the amount of absorbed light, reprocessed in IR.  The best fit is found with C00  for UDF4 ($\chi^2_{min}=42.7$,  $\rm \Delta BIC=$  23.3 and 29.81 for LF17 and CF00 respectively): the fit with C00 is better in the  NIR which is underestimated with CF00 and LF17. Here,   GS6 is best fitted with LF17 ($\chi^2_{min}=54.06$,  $\rm \Delta BIC=$  17.8 and 117 for CF00 and C00 respectively) and C00 and, to a lesser extent CF00, produce an overly high attenuation in the UV, a grayer attenuation law as LF17 better reproduces the UV with a substantial attenuation  in the  NIR.}
\label{fits}
\end{figure*}
We now fit the full  datasets with the same models and parameters defined in Section \ref{sec:method}.
First, as for the stellar continuum, we test the requirement to introduce a recent burst by calculating the BIC difference between the scenarios with and without a burst. We apply the same criterion: there is strong evidence for a burst when $\rm \Delta BIC<-6$. Using C00 and CF00 we find strong evidence for a burst for respectively 13 and 11 of the 17 galaxies. We find $\rm \Delta BIC\simeq 6$ (i.e., close to strong evidence for no burst) for UDF13 with both C00 and CF00 and also for UDF7 and GS4 with CF00. With LF17 we find strong evidence for a burst for only three galaxies (UDF3, GS3, and GS6). For all the other cases, $\rm \Delta BIC\simeq 6$. The requirement for a burst for C00 and CF00 is essentially due to the fact that the delayed SFH model alone is most of the time unable to produce enough IR emission. With LF17, the attenuation law is much grayer and the amount of attenuation is high enough to produce the dust emission observed without a burst, or with a burst of very low amplitude. \\
There is no strong evidence in favor of a simple delayed SFH without a burst. As expected when $\rm \Delta BIC\simeq 6$, the burst fractions returned by the code are always very small and never higher than $1\%$. Therefore for the sake of simplicity we adopt a delayed SFH on which a recent burst is superimposed for all the targets and dust attenuation laws. The average (median) values of the burst fraction found for C00, CF00, and LF17 are respectively 0.18 (0.20), 0.10 (0.10), and 0.03 ($10^{-6}$).

We plot in Fig.~\ref{chi2-ir} the best $\chi^2$ for the three attenuation laws.
This time, no model appears to be clearly better than the other ones for the entire sample. In Table~\ref{attlaws} we report the best attenuation law for each target based on the best $\chi^2$.
\begin{table}
\scriptsize
\begin{tabular}{l r l r rr}
\hline\hline
ID & CLS ID & Redshift & Best law & $\chi^2_{\rm min}$&Other laws \\
\hline
GS01*& 3280& 2.191&C00 &21.0& LF17\\
GS02 & 5339 & 2.326&CF00&49.0&LF17\\
GS03* &2619& 2.241& LF17&15.7&CF00\\
GS04 & 7184&1.956& CF00&37.2& LF17\\
GS05*& 9834&2.576&LF17&43.3&\\
GS06*& 14876&2.309&LF17&54.1& \\
GS07* &8409& 1.619& CF00&52.9 &\\
UDF1* &15669&2.698* &C00&46.2& \\
UDF2 & 15639&2.696*&C00&45.9 &\\
UDF3* & 15876&2.543&LF17&98.9& \\
UDF4 & 15844&2.43& C00&42.7& \\
UDF5 & 13508&1.759& CF00&41.7 &LF17/C00\\
UDF7* & 15381&2.59& C00&25.9& \\
UDF10 &12743& 2.086**&CF00 & 18.3&LF17 \\
UDF11* &12624&1.996& CF00& 34.2&LF17/C00 \\
UDF13* & 15432&2.497&C00&20.4&LF17 \\
UDF15 & 15015& 1.721**&C00&29.0&LF17 \\
\hline\hline
\end{tabular}
\caption{List of the targets with the best attenuation laws used to fit the full SED. Column 1: Target name, sources with an X-ray detection are indicated with a star. Column 2: CANDELS identifier from \citet{Guo13}. Column 3:  redshifts  from \citet{Elbaz18} except for sources with one or two stars after the redshift value (from \citet{Rujopakarn19} and \citet{Dunlop17} respectively). Columns 4 and 5: best attenuation law (column 4)  from the best $\chi^2$ (column 5). Column 6: other attenuation law(s) corresponding to $\rm \Delta BIC$ lower than 6 as compared to the best case (ordered by increasing $\rm \Delta BIC$).}
\label{attlaws}
\end{table}
The other acceptable laws on the basis of a BIC difference lower than six are also listed. The C00, CF00, and BLF7 models correspond to the best fit for seven, six, and four objects, and are acceptable  for two, one, and eight other cases respectively. In Fig.\ref{fits} we show two examples (UDF4 and GS6) for which different laws are  preferred.\footnote{The  fits corresponding to different laws often appear similar by eye, despite significant $\rm \Delta BIC$. It is due to the very high signal to noise ratio for HST, ISAAC and $\sl Spitzer$ data, which are much higher than 10 for almost all the sources. This leads to high $\chi^2$ values and large $\rm \Delta BIC$, sufficient to distinguish between different attenuation laws with this criterium. The cases presented in Fig. \ref{fits} are sufficiently different to be distinguished by eye.}

The AGN fractions (defined as the fraction of the total IR luminosity coming from the AGN) obtained for X-ray detected objects  are small ($<f_{\rm AGN}>= 0.05\pm 0.07)$ and do not have a significant impact on the results.
In the following we use the best attenuation law for each target according to Table \ref{attlaws}. The  best fits for the full SED of the whole galaxy sample  are shown in Appendix A.

\section{Dust emission\label{sec:dust}}
\begin{figure}
\includegraphics[width=\columnwidth] {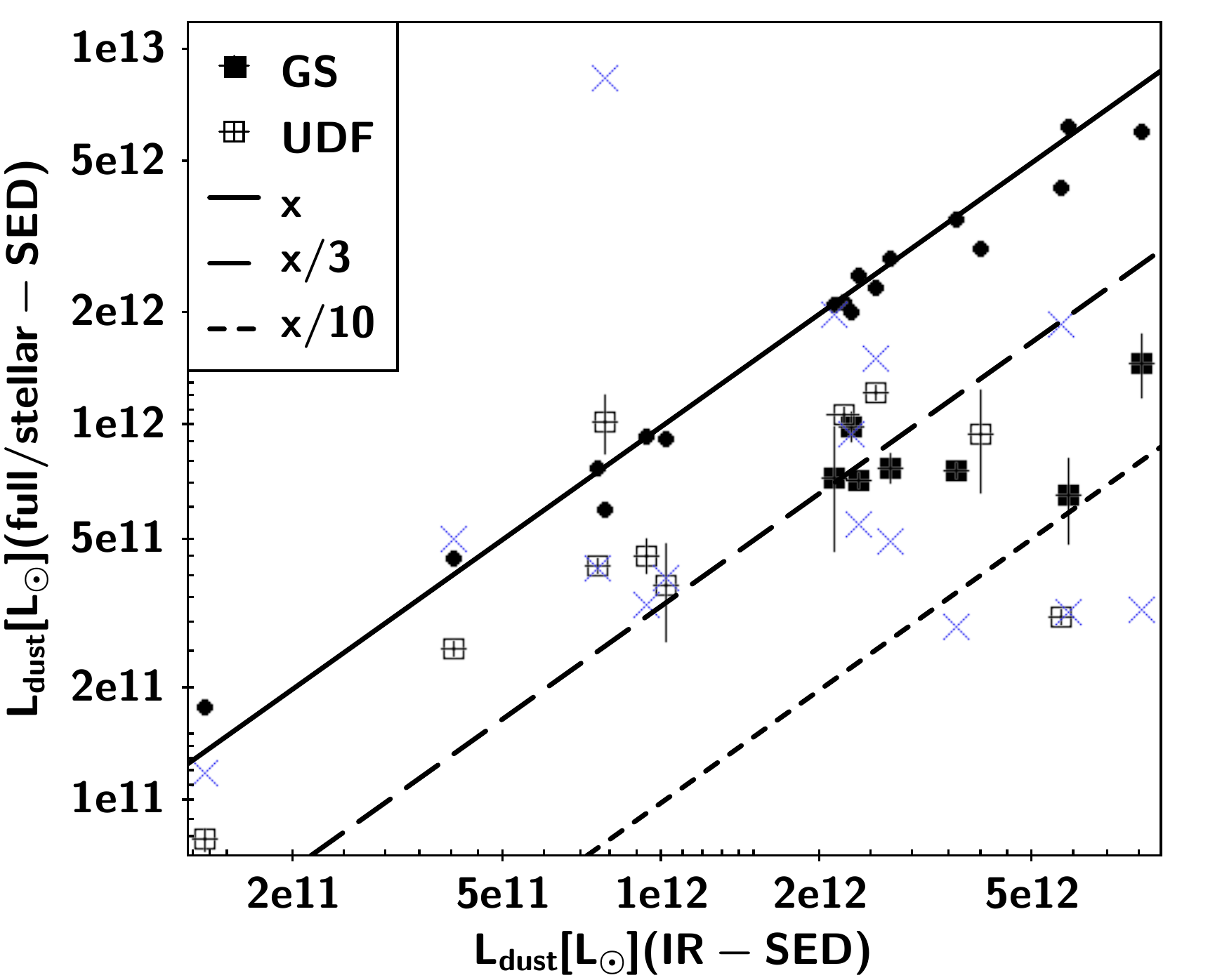}
\caption{Comparison between $\rm L_{dust}$ measured with the fit of the IR data only (MIPS to ALMA, x-axis),  with the fit of  the full SED (dots, y-axis), and with the fit of  the stellar continuum only (filled squares for GS galaxies and empty squares for UDF galaxies, y-axis). The 1$\sigma$ errors  given by CIGALE for the fit of the stellar continuum are plotted as vertical and horizontal lines. The blue crosses correspond to  $\rm L_{dust}$  values obtained with the slope of the UV continuum. The solid line represents equal quantities on both axes, and dashed and dotted lines represent a $\rm L_{dust}$ respectively   three and ten times lower on the y-axis.}
\label{Ldust}
\end{figure}

 $\rm L_{dust}$ is an output of the three  fits performed in Section \ref{sec:fit}. Here we compare the different estimates.
CIGALE outputs  $\rm L_{dust}$ due to absorption of starlight, as the result of the energy budget.  We correct the measure of  dust emission with IR data (Section 4.2) for the AGN contribution estimated with the full SED in order to compare the three measurements of $\rm L_{dust}$. While we expect a large difference between the results from the fits of the stellar continuum and the other measurements, the $\rm L_{dust}$ values measured from IR data  and from the full SED should be similar. We show in Fig.~\ref{Ldust} that the agreement between both measurements is indeed very good.

\subsection {Dust emission from the fit of the stellar continuum}
We find a substantial attenuation when fitting the stellar continuum alone with an average value of $\left<A_{\rm V}\right>= 1.6 \pm 0.5$ mag. In Fig.~\ref{Ldust} we compare  $\rm L_{dust}$ induced by this attenuation to the values   measured with IR data only in Section 4.2.  $\rm L_{dust}$ values obtained with the fit of the stellar continuum are found to be lower than the luminosities measured with IR data. The median value of the ratio between both measurements is $0.41$ for the whole sample, and $0.56$ and $0.28$ for the UDF and GS galaxies respectively. In one case (UDF13) the IR emission is over-predicted from the fit of the stellar continuum alone, with an excess of 30$\%$. In section \ref{sec:fit} we show that both the stellar and full SEDs of UDF13 can be fitted without adding a burst, for any attenuation law. No extra attenuation is needed to explain  $\rm L_{dust}$ measured from {\sl Herschel} and ALMA data. Conversely, for two galaxies (UDF3 and GS6), less than $10\%$ of the dust emission can be explained from the rest-frame UV to NIR fit. We found that a burst must be added to fit the full SED of these galaxies with any of our attenuation laws. We see in Section \ref{sec:sfr-mstar} that both are the most active galaxies of our sample in star formation.

Both \citet{Dunlop17} and \citet{Elbaz18} also report  a substantial attenuation from the analysis of the stellar continuum. This result is puzzling: in these objects  the ALMA emission does not match the UV rest-frame emission and in  some cases there is  even a full disconnection between the two. 
Our fits are averaged over the whole of the galaxy in each case and are not necessarily valid for the specific regions where the UV emission is detected. The light distribution in the HST/ACS F160W filter (corresponding to visible rest frame) extends much farther than what is seen in the HST/ACS F606W and F814W filters, which correspond to rest-frame UV \citep{Rujopakarn16,Elbaz18}.  These different spatial distributions  may be  of consequence for energy balance models even when only the stellar continuum is considered. One way to check the energetic  link between the UV and dust  emissions  is to estimate $\rm L_{dust}$ only from the rest-frame UV  emission as we do in the following section.

\subsection{Dust emission from the slope of the UV continuum}
\begin{figure}
\begin{center}
\includegraphics[width=7cm] {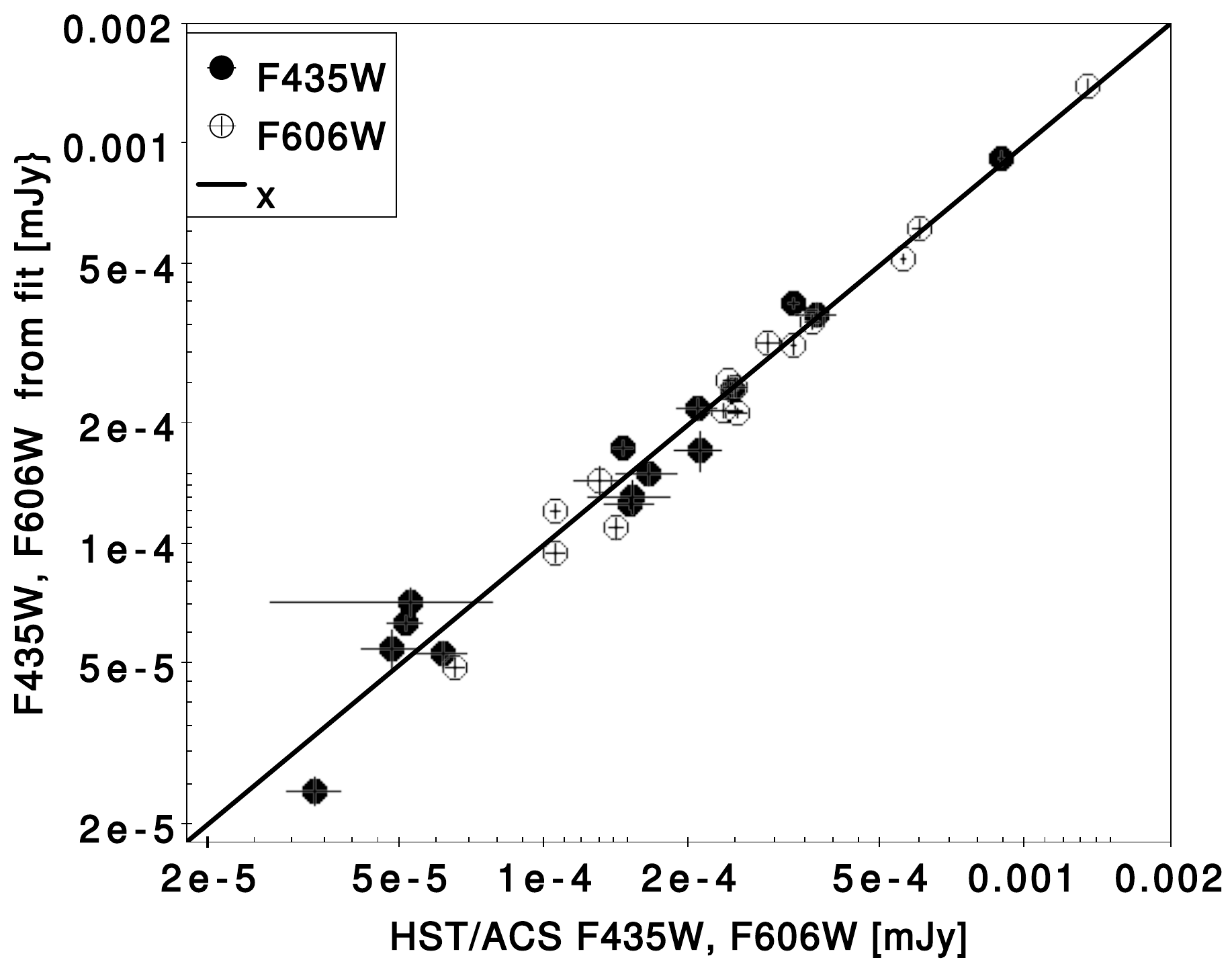}
\end{center}
\caption{Comparison between the observed fluxes (x-axis) sampling the UV rest-frame continuum, HST/ACS 435W (filled circles) and 606W (empty circles), and the estimations from the fits (y-axis). The 1$\sigma$ errors are indicated as well as the 1:1 relation (solid line).}
\label{Fuv}
\end{figure}

At the median redshift of the sample ($z=2.3$) the HST/ACS F432W and F606W filters correspond to 130 and 180~nm respectively, allowing us to estimate the slope of the UV continuum, which is expected to follow a power law ($f_{\lambda} \propto \lambda^{\beta}$). We exclude the most distant galaxies of the sample from this analysis, namely UDF1 and UDF2. At $z\simeq2.7$, the F435W filter samples the emission below 120~nm where the attenuation curve is not well constrained and is likely to depart from the Calzetti law \citep{Leitherer02, Buat02,Reddy16}. We calculate $\beta$\footnote{$\beta= -6.9\times \log(f_{F435W}/f_{F606W})-2$ with fluxes expressed in mJy} using the fits of the stellar continuum alone after ensuring that the Bayesian estimates of the fluxes in the F435W and F606W filters are in very good agreement with the observed values (Fig.~\ref{Fuv}). With $\beta$ at hand, we estimate the IRX ratio using the IRX-$\beta$ relation valid for local starburst galaxies. We use the inner relation of \citet{Overzier11}, close to the original fit of \citet{Meurer99}. $\rm L_{dust}$ is then deduced by multiplying IRX with the UV rest-frame luminosity estimated by CIGALE and corresponding to the GALEX/FUV filter.

We plot the $\rm L_{dust}$ obtained from IRX-$\beta$ in Fig.~\ref{Ldust} for comparison with the results obtained in the previous subsection. We find similar trends considering either the fit of the stellar continuum or using only $\beta$. Given the large uncertainties on the measure of $\beta$ with only 2 bands and on the validity of the IRX-$\beta$ relation for these objects, we do not consider the values based on the IRX-$\beta$ relation as sufficiently reliable. We only use them confirm the rather large attenuation found for the UV emission, which corresponds to a substantial fraction of the total dust emission of these galaxies, at least in terms of energy. \citet{GomezGuijarro18} also found that the dust emission is always associated with the reddest stellar component.

In the case of UDF13, $\rm L_{dust}$ calculated from $\beta$ is much higher than the total observed dust luminosity from the dust only model. Similarly, $\rm L_{dust}$ measured from its stellar continuum is also larger than when measured from the dust emission only, although with a much lower excess. This galaxy is not starbursting as we showed before, and we cannot interpret its relatively red UV slope ($\beta = 1.2$) as being exclusively due to dust attenuation. 

\begin{figure}
\includegraphics[width=\columnwidth] {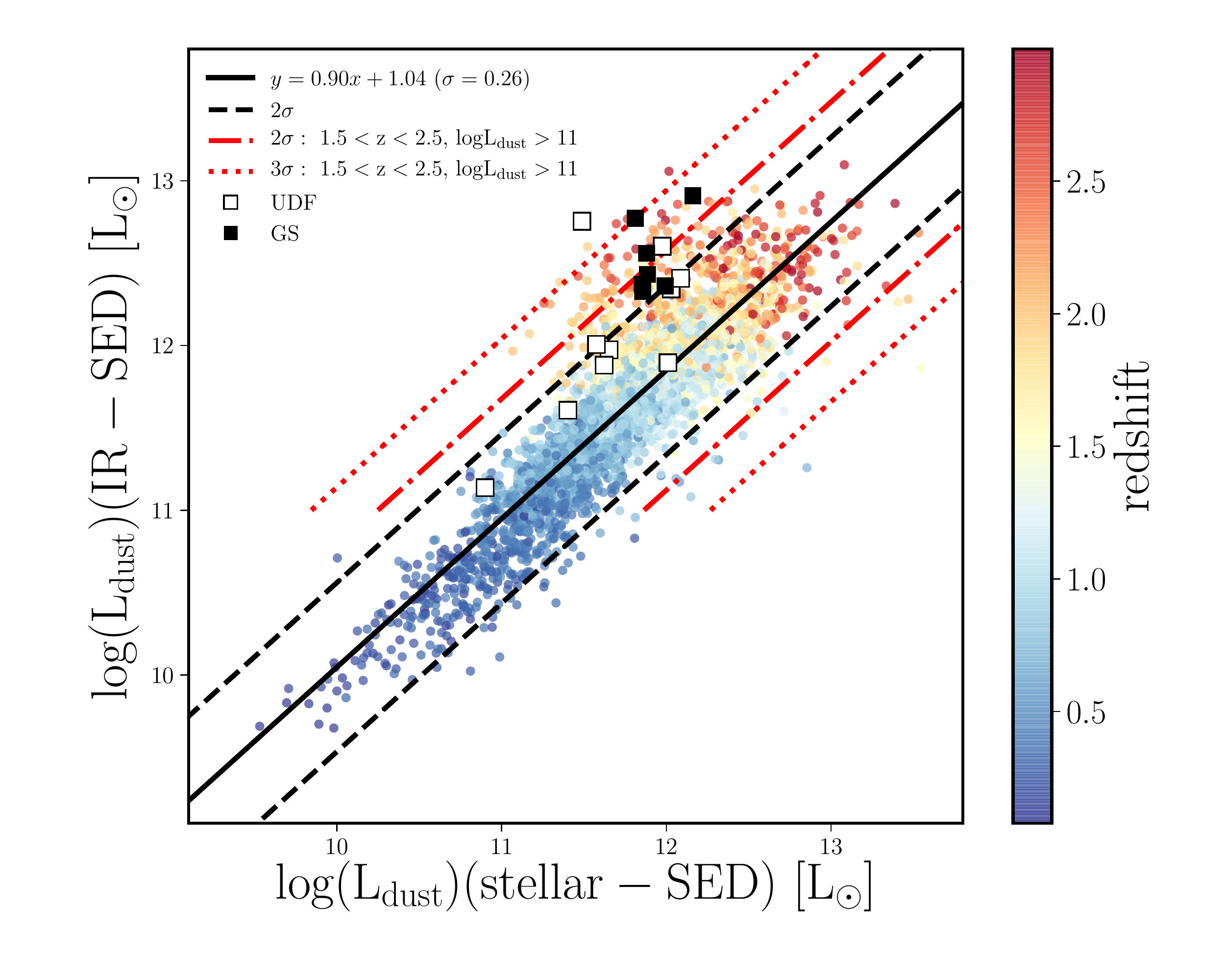}
\caption{ Comparison between $\rm L_{\rm dust}$ estimated  from the fit of the stellar continuum (x-axis) and with IR data (y-axis). Filled squares are used to represent GS galaxies  and UDF galaxies are plotted  with empty squares. The COSMOS sample is plotted with dots color coded with redshift.  The black solid and dashed lines represent the  linear regression and 2$\sigma$ dispersion  for the full COSMOS sample. The red dot-dashed and dotted line the 2 and 3 $\sigma$ dispersion for the COSMOS galaxies with $1.5<z<2.5$ and $L_{\rm dust} >10^{11}~L_{\odot}$.}
\label{Ldust-comp}
\end{figure} 

\subsection{Comparison of the HUDF, GOODS-S and COSMOS/HELP surveys}

As part of the HELP project\footnote{ The Herschel Extragalactic Legacy  Project (https://herschel.sussex.ac.uk/). HELP data products are distributed on HeDaM  (http://hedam.lam.fr/HELP/ )}, \citet{Malek18} analyzed the SED of more than 40 000 galaxies of the pilot field ELAIS-N1 and performed SED fitting with CIGALE. They  estimated $\rm L_{dust}$  from the stellar continuum for  686 galaxies in the ELAIS-N1 field detected with {\it Herschel} with a signal-to-noise ratio (S/N) larger than two. They found a very good agreement and a tight relation between both estimations (their figure 12.c). 

The ELAIS-N1/HELP field is not very deep with a mean redshift of 0.93 and not suitable for a comparison with our current study. In order to compare samples in the same range of redshift and $\rm L_{dust}$, we  consider  the COSMOS/HELP field for which similar data products are available.  As for the ELAIS-N1 study, we select galaxies with {\sl Herschel}  detections (at least two  SPIRE fluxes with an SNR>3 and at least one PACS flux with SNR > 2), at least six photometric measurements of the stellar continuum, and good-quality fits of the stellar continuum and IR SED (for more details of the selection see \citet{Malek18}).  We select 3365 galaxies  with $z<3$ and 694 with $1.5<z<2.5$. 
In Fig.\ref{Ldust-comp} we plot the two estimations of $\rm L_{dust}$  (from the stellar continuum alone and with IR data)  for  the COSMOS sample, the correlation is good  with a correlation coefficient  of 0.90.  We can see on the plot that the dispersion around the linear fit increases with redshift and luminosity. The dispersion  for the full sample  is $\sigma=0.26$ and increases to $\sigma=0.36$ for the subsample of  galaxies with $1.5<z<2.5$ and $ L_{\rm dust} >10^{11}~L_{\odot}$ . The values for the GS and UDF galaxies are over-plotted. All of them lie on or  above the linear regression line found for the COSMOS sample. All but one of the UDF galaxies (UDF3) lie within the $2 \sigma$ limit for $1.5<z<2.5$ and  $ L_{\rm dust}> 10^{11}~L_{\odot}$. The GS galaxies are  close to  or   above this 2$\sigma$  limit   but they remain below $3\sigma$. This difference of behavior between UDF and GS objects is expected since HUDF is a   blind survey whereas  the sources of \citet{Elbaz18} were selected to be very massive and IR bright. We return to this comparison in Sect. 6.3 after SFR and stellar masses estimations and in relation to the position of the galaxies in the MS. Nevertheless,  a full comparison between ALMA- and {\sl Herschel}-selected samples  is beyond the scope of this paper, and would include a careful analysis of the selection bias for each sample. Multiwavelength analyses, including SED fitting,  of larger ALMA blind surveys like the GOODS-ALMA and ASAGAO surveys \citep{Franco18, Hatsukade18}  are also needed to perform a robust statistical comparison with the large {\sl Herschel} surveys.

\section {Star formation rate and stellar mass determinations\label{sec:sfr-mstar}}
\begin{table*}
\centering
\begin{tabular}{ccccc}
\hline\hline
  ID &$\rm L_{dust}$ (IR SED) &SFR & $\rm M_{\star}$ (full SED) &$\rm M_{\star}$ (stell. SED) \\
   &$\rm 10^{12} L_{\odot}$ & $\rm M_{\odot} ~yr^{-1}$ & $\rm 10^{10} M_{\odot}$&$\rm 10^{10} M_{\odot}$\\
\hline
  GS1   &  2.30$\pm$ 0.11& 218$\pm$ 14 & 5.56$\pm$ 1.03& 7.18$\pm$ 0.83\\
  GS2   &  2.35 $\pm$  0.12& 254 $\pm$ 13 & 9.13 $\pm$ 1.4 & 5.89 $\pm$ 0.43\\
  GS3   & 3.63 $\pm$  0.18& 323 $\pm$ 16 & 19.53 $\pm$ 0.98 & 8.49$\pm$ 1.49\\
  GS4   &  2.13 $\pm$  0.10 & 154 $\pm$9 & 21.08 $\pm$ 2.62 & 10.66 $\pm$ 2.35\\
  GS5   &  8.10$\pm$  0.40 & 432 $\pm$ 22 & 58.78 $\pm$ 2.93 & 19.61$\pm$ 0.98\\
  GS6   &  5.91 $\pm$ 0.30 & 567$\pm$ 28& 33.90 $\pm$ 1.95 & 8.83 $\pm$ 0.60\\
  GS7   &  2.70$\pm$ 0.13 & 274 $\pm$14 & 14.42 $\pm$ 2.30 & 8.80 $\pm$ 1.21\\
  UDF1  &  4.00 $\pm$  0.20 & 334$\pm$ 17 & 5.00 $\pm$ 0.82 & 7.34 $\pm$ 0.43\\
  UDF2  & 2.21 $\pm$  0.15& 229$\pm$ 15 & 6.91 $\pm$ 1.02 & 8.30$\pm$ 0.69\\
  UDF3  &  5.68 $\pm$  0.28 & 491 $\pm$ 25 & 6.35 $\pm$ 0.32 & 2.89 $\pm$ 0.29\\
  UDF4  &  0.94 $\pm$  0.048 & 104$\pm$ 7 & 2.24 $\pm$ 0.47 & 3.36 $\pm$ 0.31\\
  UDF5  &  0.76 $\pm$  0.04 & 72 $\pm$ 5 & 4.50 $\pm$ 0.22 & 3.31 $\pm$ 0.32\\
  UDF7  &  1.05 $\pm$ 0.05 & 95 $\pm$ 13 & 3.18 $\pm$ 0.48 & 4.69 $\pm$ 0.84\\
  UDF10 &  0.40 $\pm$  0.02 & 45 $\pm$ 5 & 2.32 $\pm$0.25& 1.98 $\pm$ 0.19\\
  UDF11 &  2.56 $\pm$  0.13 & 245$\pm$ 12 & 13.48$\pm$ 1.79& 9.05 $\pm$ 1.03\\
  UDF13 &  0.79 $\pm$  0.04 & 46 $\pm$ 8 & 7.52 $\pm$ 0.37 & 6.99 $\pm$ 0.67\\
  UDF15 &  0.14 $\pm$  0.01 & 20 $\pm$ 3 & 0.89$\pm$ 0.16 & 1.15$\pm$ 0.16\\
\hline\hline

\end{tabular}
\caption{Physical quantities estimated from the fits performed in section 4 with the 1$\sigma$ error given by CIGALE. Column 1: Target names  of the sources. Column 2: $\rm L_{dust}$ from the fit of the IR data (section 4.2). Columns 3 and 4: SFR and stellar masses from the fit of the full SED (section 4.3). Column 5: stellar masses from the fit of the stellar continuum alone (section 4.1).}
\label{bayesparam}
\end{table*}

The main parameters extracted from the SED fitting are usually stellar masses and SFRs.  We list in  Table \ref{bayesparam} the stellar masses obtained from the fit of the stellar continuum  together with the SFR and stellar masses obtained using the full SED. We discuss their reliability here.

\subsection{Stellar mass}
\begin{figure}
\includegraphics[width=8cm] {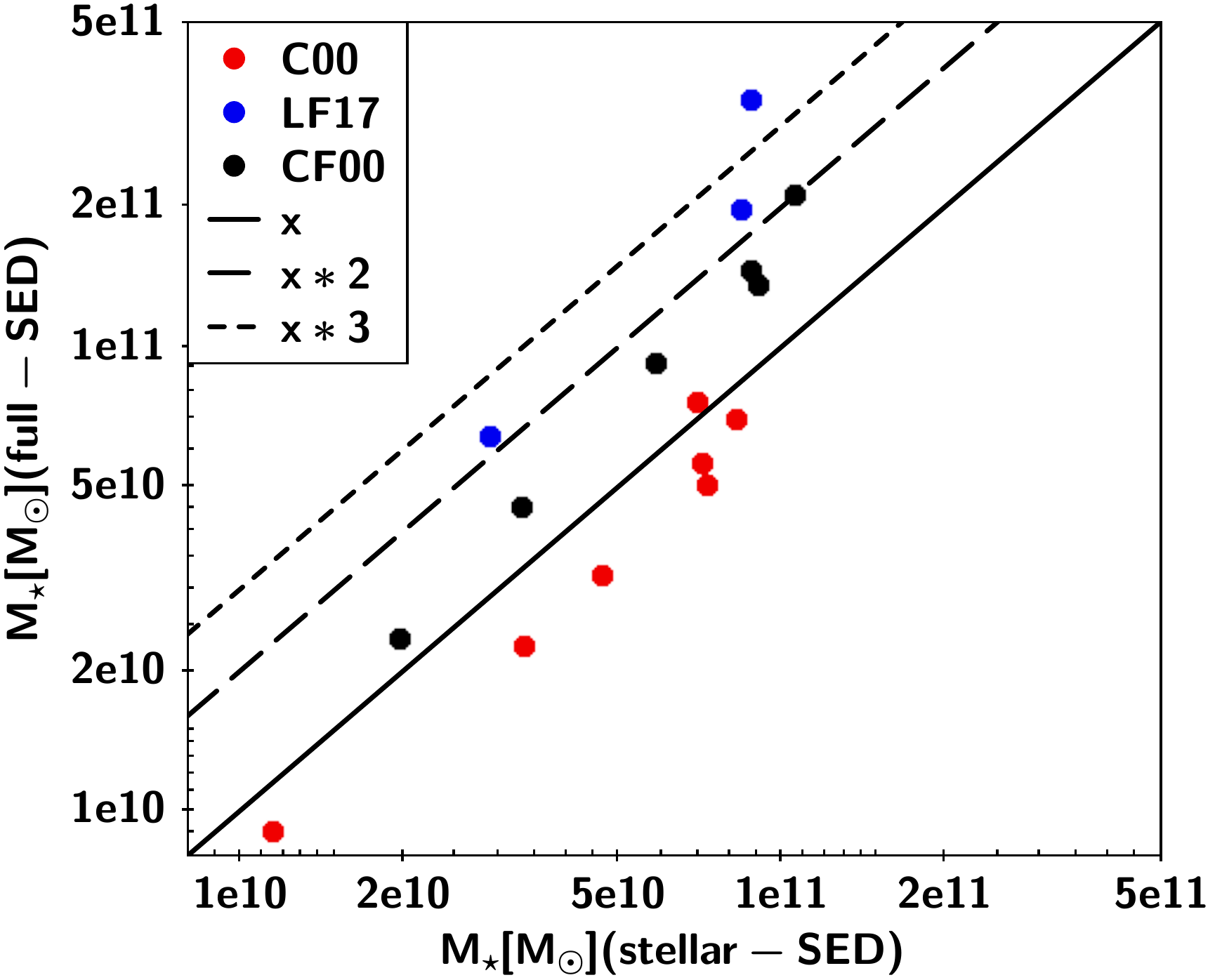}
\caption{Comparison between stellar masses ($\rm M_\star$) measured with the fit of the stellar continuum and C00 (x-axis) and the fit of the full SED with the best attenuation law (y-axis). The different attenuation laws adopted for the fit of the full SED are indicated with red (C00), blue (LF17), and black (CF00) filled circles. The solid line represents the 1:1 relation, the dashed (resp. dotted) line a stellar mass two (resp. three) times lower on the y-axis.}
\label{mstar}
\end{figure}

In Fig.~\ref{mstar} we compare the stellar masses measured from the  stellar continuum alone (with C00 and a simple delayed SFH) and the full SED (with the best attenuation law and a delayed SFH with a recent burst). As expected the choice of the attenuation law has a strong impact on the stellar mass.
The flatter attenuation laws are always favored for the most massive systems ($> \sim 10^{11} M_{\sun}$). There is a good agreement between the masses when C00 is used for both fits of the full SED and stellar continuum. In this case, we obtain only slightly lower values with the full SED (factor of 0.8 on average). We attribute this difference to the presence of a burst of young stars in the case of the full SED fitting,  which  over-shines older stellar populations \citep{Pforr12}. We find higher stellar masses with CF00 and LF17 with an average systematic difference of close to a factor of two (reaching 2.8 for LF17 against 1.5 for CF00) in agreement with the findings of \citet{Malek18}. This is likely due to the substantial attenuation affecting rest-frame stellar emission and even the NIR with CF00 and LF17.

The stellar masses obtained by \citet{Dunlop17} from their fit of the UV-to-NIR SED of UDF galaxies agree within 0.15 dex with our estimations from the stellar continuum. This agreement is expected since both fits are performed assuming C00.
 
\subsection{Star formation rate}
\begin{figure}
\includegraphics[width=8cm] {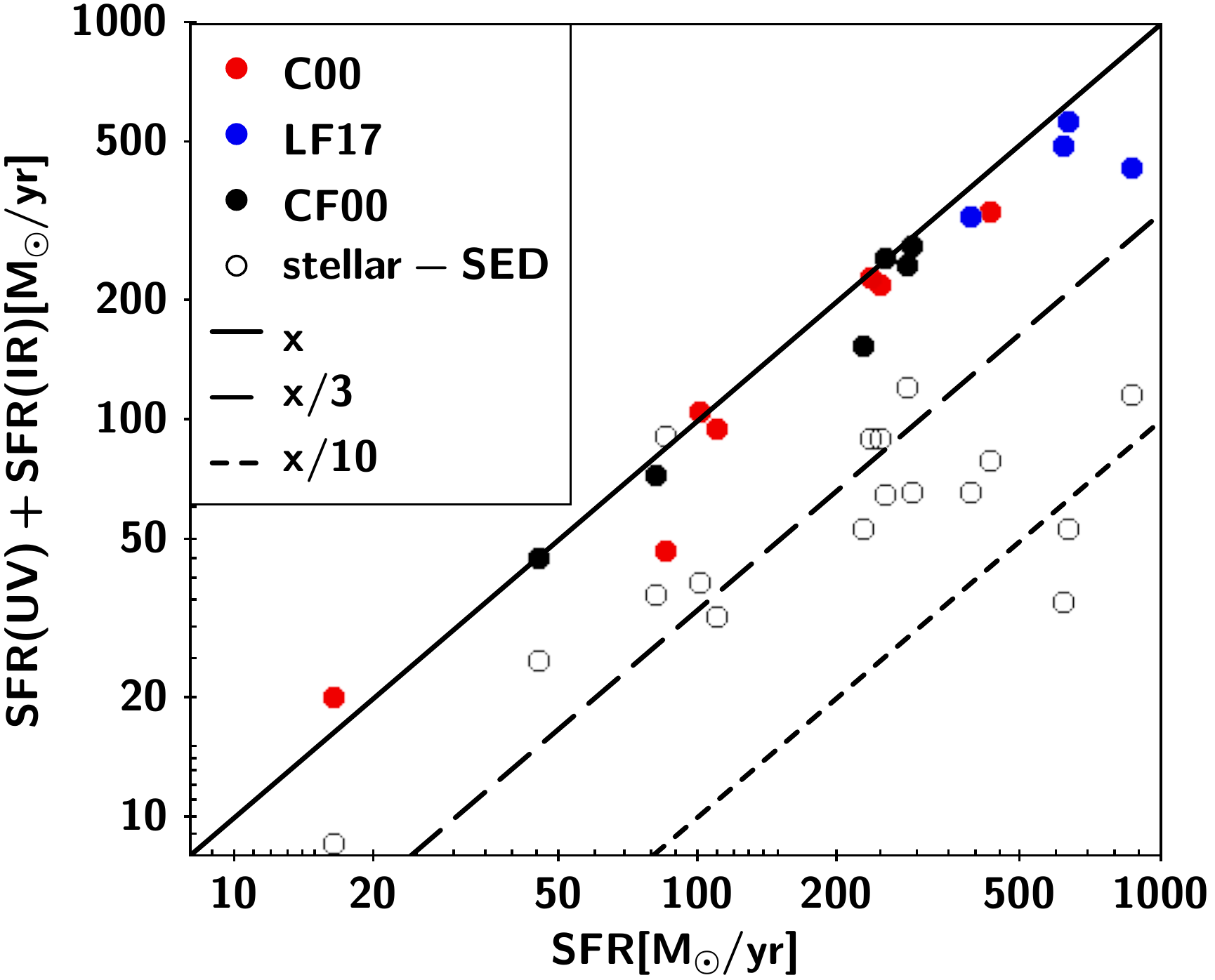}
\caption{Comparison between the SFRs measured from SED fitting (x-axis) and the sum of SFR(IR) and SFR(UV), (y-axis). The filled circles correspond to the fit of the full SED as in Fig. \ref{mstar}. The empty circles correspond to the SFRs obtained with the fit of the stellar continuum alone. The solid line represents equal quantities on both axes; the dashed and dotted lines represent a quantity  respectively   3 and 10 times lower on the y-axis.}
\label{sfr}
\end{figure}

In Fig.~\ref{sfr}, we compare the SFR measured from the full SED with the simple approach of adding the SFR obtained from the total dust and UV emissions: SFR(IR)+SFR(UV).
 We adopt here the calibrations of \citet{Buat08} between luminosities and SFR. We use   $\rm L_{dust}$ values obtained with the fit of the IR data  and corrected for the AGN contribution as explained in Section 5.  The SFRs(IR)  agree within a factor two with the same quantity measured by \citet{Dunlop17} for the UDF galaxies ($\rm SFR_{FIR1}$ in their Table 4) except for UDF3 for which our measurement is 2.5 higher than the value they reported. The rest-frame UV luminosity (without dust correction) is an output of CIGALE already used in section 5.2: it comes from the fit of the stellar continuum alone. 

 We find a satisfying agreement between the two estimates of the SFRs. The SFR(IR)+SFR(UV) calculation is based on very simple assumptions on the SFH and dust heating. We therefore do not expect a perfect agreement with our measurements coming from SED fitting. In Fig.~\ref{sfr} we also plot the SFRs measured from the stellar continuum alone. We find they are much lower than the ones based on the full SEDs by a factor similar to the one found for the estimation of  $\rm L_{dust}$.

Once again, UDF13 departs from the general trend. Its SFR measured from the full SED is lower than both the addition of the IR and UV SFRs and the SFR measured from the stellar continuum.The difference with the former comes from the SFH used for the fit (a delayed SFH with a peak 1 Gyr before and no recent burst with a burst fraction equal to $10^{-7}$) versus the simple assumption of a constant SFH over 100 Myr to calculate SFR(IR)+SFR(UV) \citep[e.g.,][]{Kennicutt12,Buat08}.  The discrepancy with the latter is explained by  higher $\rm L_{dust}$ inferred from the stellar continuum alone. 

\subsection{Position on the main sequence of star forming galaxies}
\begin{figure}
\includegraphics[width=\columnwidth] {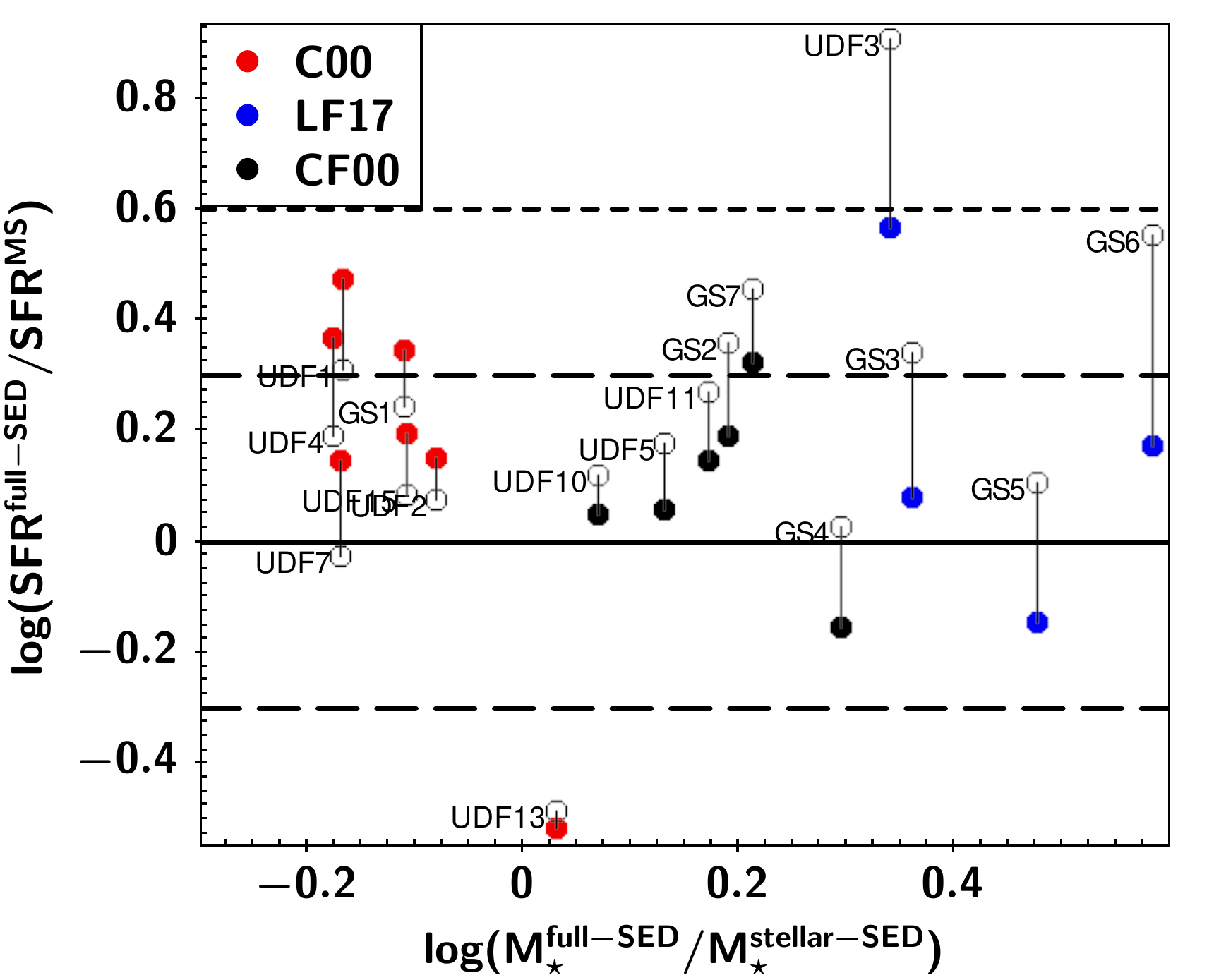}
\caption{Relative position of the galaxies on the MS. The ratio of stellar masse ($\rm M_\star$) measured with the fit of the full SED and of the stellar continuum is shown on the x-axis. The y-axis shows the  ratio of  SFR measured with the fit of  the full SED and  SFR calculated on the MS. The latter is calculated with the \citet{Schreiber15} relations using either the stellar mass measured  with the fit of the stellar continuum only (empty circles) or with the fit of the full SED (filled circles, red for C00, blue for LF17, and black for CF00). Vertical lines connect the two SFR ratios calculated with the different stellar mass estimations for each galaxy.}
\label{MS}
\end{figure}

We define the MS with the relations of \citet{Schreiber15} as a function of stellar mass and redshift. For each source we consider the two stellar masses previously computed either from the stellar continuum or from the full SED and we calculate the two SFRs corresponding to these stellar masses for a galaxy   on the MS.  In Fig.~\ref{MS}, for each target, we  compare these two SFRs with the SFR measured previously from the fit of the full SED and plot the ratio of these values as a function of the ratio of the two stellar masses.
We assume a dispersion of 0.3 dex for the MS \citep{Schreiber15}. With the stellar masses measured with the stellar continuum, six galaxies are found above the MS and one below. If we define a starburst as a galaxy with an SFR exceeding the MS by a factor at least four, only one galaxy is found to be starbursting (UDF3). Another one (GS6) has an SFR 3.5 times higher than the MS. If we now use the stellar masses measured from the full SED, five galaxies are found above the MS but close to the boundaries of the relation, and no galaxy is starbursting. UDF13 is still well below the MS and UDF3 very close to starbursting.

 \citet{Dunlop17}  concluded that their data were completely consistent with the published MS relations at $z\simeq 2$.  \citet{Elbaz18}  also introduced the variation of  the MS with redshift by using the relations of \citet{Schreiber15} as we did above. They  also found  that  the  UDF and GS galaxies are located in the upper part of the MS and only  UDF3, GS6, and GS7 are starbursting, with a SFR more than four times above the MS.  In our analysis, GS6 and GS7 appear less extreme but still well above the MS for a stellar mass estimated with the stellar continuum.
 \begin{figure}
\begin{center}
\includegraphics[width=\columnwidth] {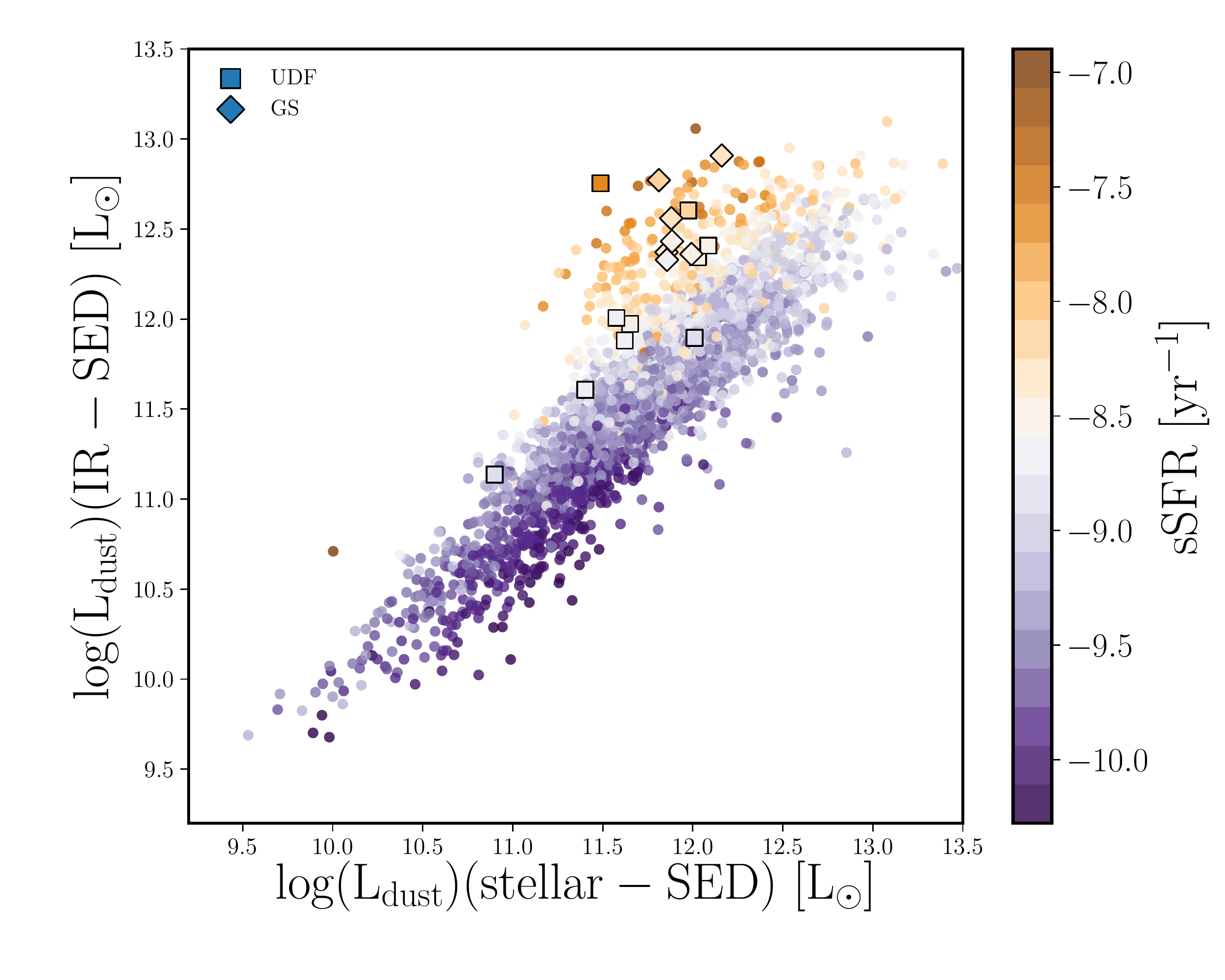}
\end{center}
\caption{ Comparison between $\rm L_{\rm dust}$ estimated  from the fit of the stellar continuum (x-axis) and with IR data (y-axis) as in Fig. \ref{Ldust-comp}.  Diamonds  are used to represent GS galaxies  and UDF galaxies   are plotted with squares. The COSMOS sample is plotted with dots. The symbols are color coded with sSFR.}
\label{Ldust-ssfr}
\end{figure} 

The difference found between $\rm L_{dust}$ estimations with the fit of either IR data or  stellar continuum discussed in section 5.3 is found to increase with  star forming activity. In Fig.\ref{Ldust-ssfr} the same quantities as in Fig.\ref{Ldust-comp} are plotted but this time color coded with the specific SFR defined as the SFR  from the fit of the full SED  divided by the stellar mass from the fit of the stellar continuum. For   COSMOS, UDF, and GS galaxies the excess of dust emission measured with IR data  increases with sSFR. The fact that all our UDF and GS galaxies are found in the upper part of the MS explains why they depart from the  mean trend found by \citet{Malek18}  between both estimations of $\rm L_{dust}$ for the ELAIS-N1 field and confirmed here in the COSMOS field, but all  remaining  with the  3$\sigma$ limit.

\section{Dust attenuation laws and compactness\label{sec:laws}}

\subsection{ALMA, H-band radii, and attenuation laws}
\begin{figure}
\begin{center}
\includegraphics[width=6cm] {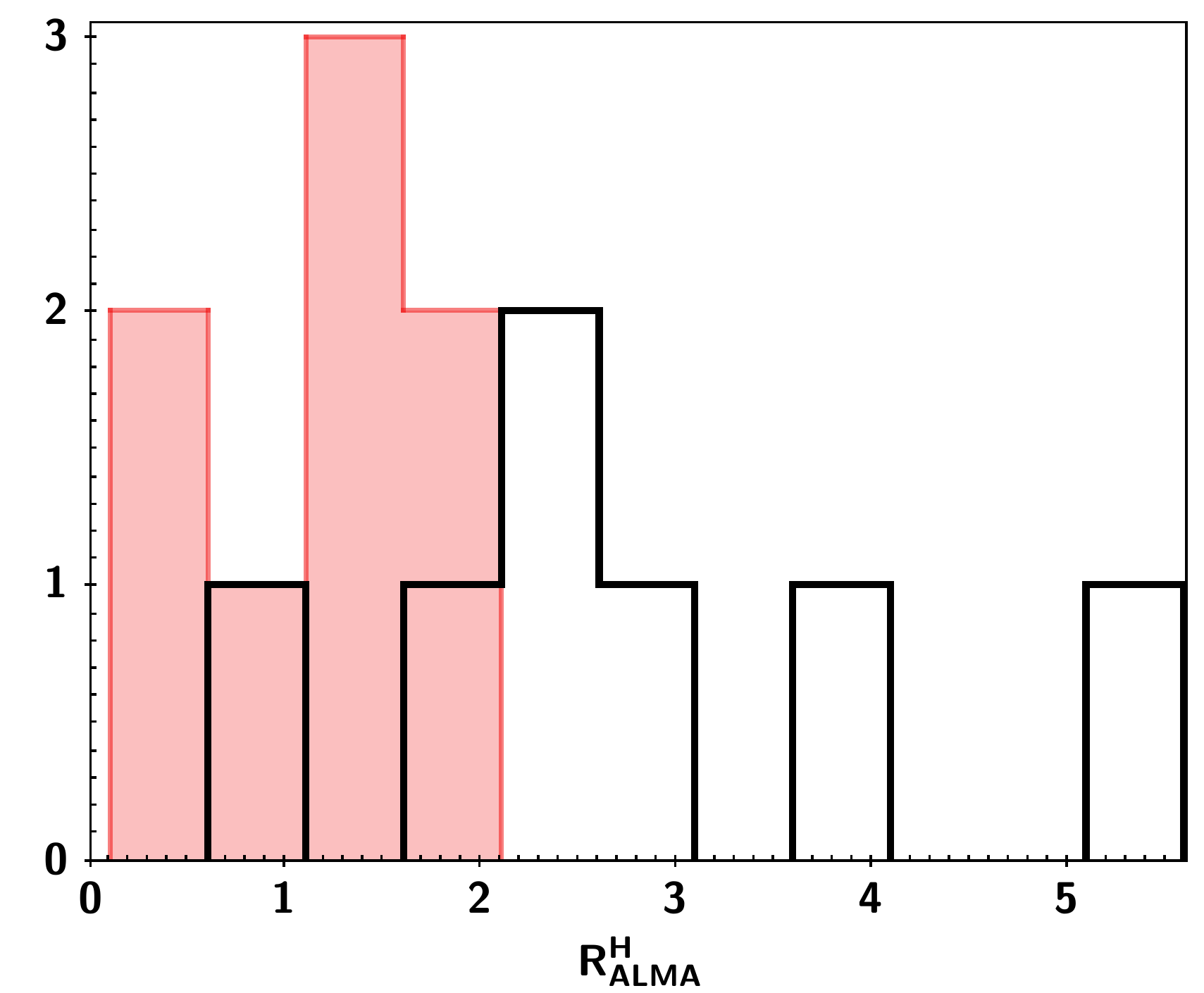}
\end{center}
\caption{Histogram of the ratio of the H-band radius to the ALMA radius, $\rm R^H_{ALMA}$. The filled histogram represents  galaxies for which C00 gives a satisfying fit, and the empty histogram represents galaxies for which a shallower power-law attenuation (CF00 or LF17) is needed.}
\label{radii}
\end{figure}

The compactness of the most intense sources detected by ALMA leads us to compare the effective radii measured in the H and ALMA bands respectively. We use the circularized half-light radii of \citet{Elbaz18} and we define their ratio: $\rm R^H_{ALMA} = R(H)/R(ALMA)$. \citet{Rujopakarn16} also measured radii corresponding to the star formation and stellar mass distributions for the UDF galaxies. Although these radii correspond to the distribution of physical quantities that we are discussing in this paper, we prefer to adopt the \citet{Elbaz18} radii values to have homogeneous measurements for both GS and UDF galaxies. The radii measured by \citet{Rujopakarn16} for the UDF galaxies correspond to slightly smaller radius ratios, that is, more similar star formation and stellar mass distributions; using them  would not change our conclusions. We calculate $\rm R^H_{ALMA}$ for 15 of our 17 galaxies (UDF10 and UDF15 are not included in the \citet{Elbaz18} sample and are not measured by \citet{Rujopakarn16}). In Fig.~\ref{radii} we split the sample in two parts: the galaxies which can be fitted with C00 (from Table 2 with C00 in column 4 or 5)  and the galaxies which  are only well fitted with a power-law curve (CF00 or LF17). We plot the histogram of  $\rm R^H_{ALMA}$ for the two subsamples. All the galaxies which can be fitted with C00  correspond to $\rm R^H_{ALMA} <2$.  This is the case of all the UDF galaxies except for UDF3, for which $\rm R^H_{ALMA}= 1$ and the best law is LF17. Six out of the seven galaxies whose SEDs are only well fitted with a power-law curve (CF00 or LF17)  exhibit $\rm R^H_{ALMA} \ge 1.6$. All the GS galaxies are in this configuration except GS1, with $\rm R^H_{ALMA} =0.9$ and a best fit with C00. 

Our current study is limited by the small number of sources. The  need for flatter attenuation laws for sources  with a very compact dust distribution has to be confirmed with  a larger sample of sources with measured morphological parameters from optical and ALMA images \citep[e.g.,][and references therein]{Fujimoto18}. This is planned for a future study. 

\subsection{Validity of attenuation laws}
The galaxies of our sample with the largest difference between dust and stellar distributions correspond to a power-law attenuation curve, which leads to a substantial attenuation up to the rest-frame NIR. This attenuation is particularly large when LF17 is used (UDF3, GS3, GS5, and GS6, as seen in Fig. A1). The better fits obtained with a shallower attenuation law reflect the need to produce a large dust emission with a substantial to strong attenuation of the whole  stellar continuum. Variable attenuation laws from the UV to the NIR are found from dust radiative transfer calculations, either for simplified models of galactic disks or coupled with hydrodynamical simulations  and a substantial to  high optical depth experienced by all the stellar populations  leads to  a flattening of the effective attenuation law. 
\citep{Chevallard13, Trayford19}.    \citet{Roebuck19} performed radiation transfer on a library of hydrodynamic simulations for isolated disk and mergers. They  found flat attenuation curves  for the dusty phases of each simulation. These curves are well reproduced by the LF17 law and are found consistent with a heavily obscured young stellar population with a contribution of  an unattenuated older stellar  population, with a similar stellar mass in both components.  

Variations of the effective UV attenuation curve are reported by several studies based on observations and the IRX-$\beta$ scatter plot is recognized as a diagnostic for these variations \citep[e.g.,][]{Salmon16, LoFaro17, Popping17, Narayanan18, Salim19}. A flattening at longer wavelengths than the UV is much more difficult to identify on the observed SED because of the well-known degeneracy between age and attenuation. The analysis of the UV to submm SED puts some constraints on this flattening from an energy balance approach, as we did in this study and in \citet{LoFaro17}. As mentioned above, radiation transfer models link the  flattening of the effective attenuation law  to an increase of the effective optical depth at the corresponding wavelength. It is not clear if this is really valid for our targets with $\rm R^H_{ALMA} >2$ as their rest-frame visible stellar light extends well beyond the ALMA emission. In this case energy balance SED modeling may over-estimate the stellar mass by adding extra attenuation up to the NIR rest frame if a substantial part of this emission comes from unattenuated  old stars.
 
When the stellar and dust size distributions are more similar, the C00 law gives satisfying fits. The stellar masses from the full SED and the stellar continuum only are similar. We note that we cannot exclude a moderate flattening in the NIR:  the steep decrease of C00 for $\rm \lambda>0.5~\mu m$ (Fig.~\ref{dustlaw}) is not easy to reproduce, even when a single stellar population is assumed \citep{Seon16, Buat19}.  In these galaxies fitted with C00, the ALMA and HST distributions can also be very different at small scales with a smooth cold dust distribution and a patchy/disturbed stellar emission as reported by \citet{Rujopakarn19} with high ALMA spatial resolution imaging of three UDF galaxies. Nevertheless their dust emission is fitted with the ``classical'' C00.

The case of UDF3 is not consistent with the general trend. Whereas the ALMA and H-band radii are similar, LF17 fits this galaxy better.  In this case, a gray attenuation curve is likely to be representative of the stars and dust distributions inside the galaxy and the stellar mass estimated from the fit of the stellar SED with C00 underestimates the stellar mass by a factor 2.6. We note however that this is our worst fit with any attenuation law considered in this work (Table 2 and Fig. A1)

Our analysis shows that some indirect information can be retrieved from the UV to submm (i.e stellar and dust) energy balance modeling. According to Fig.~\ref{radii}, when we can  fit the full SED  with C00, a similar radius is found for the stellar population in the NIR and the cold dust detected by ALMA. Conversely, for galaxies with a very compact dust distribution, some additional obscuration is needed to produce enough dust emission. In this work we model it with a grayer attenuation law. This modification  of the attenuation law seems to  reflect  heavily obscured objects with  different dust and observed stellar  distributions, as also  identified  in the nearby universe \citep{Charmandaris04, Howell10} and  in various studies at higher redshift based on  ALMA data \citep[e.g.,][]{Hodge16, GomezGuijarro18}. The ULIRGs studied by \citet{LoFaro17} and the galaxies of the ELAIS-N1 field studied by \citet{Malek18}  and  fitted with   flat attenuation laws  could also be  in  this configuration.

 \section{Conclusions\label{sec:conclusion}}
Here, we analyze the SEDs of 17 dust-rich galaxies at $z\sim 2$ detected with ALMA from the blind survey conducted by \citet{Dunlop17} in the HUDF and the massive {\it Herschel} galaxies in the GOODS-{\it South} field by \citet{Elbaz18}. We performed several fits with CIGALE with different attenuation laws, considering only the stellar continuum, the dust emission, and the full dataset.

The fit of the stellar continuum with the starburst law of \citet{Calzetti00} reveals an obscuration which corresponds on average to one-third (UDF galaxies) to half (GS galaxies) of the total dust emission detected with {\it Herschel} and ALMA. Qualitatively, the analysis of the slope of the UV continuum confirms this obscuration of the stellar continuum. We compared our estimations of dust emission to the same quantities measured for a sample of galaxies of similar $\rm L_{dust}$ and redshift in the COSMOS field. The deficit of $\rm L_{dust}$ found from the fit of the stellar continuum is on average absent for the COSMOS galaxies, but is consistent within $2\sigma$ and $3\sigma$ respectively for the UDF and GS galaxies. In all samples, UDF, GS, and COSMOS, the deficit is found to strongly correlate with sSFR. 

The fit of the full SED (stellar and dust components) with CIGALE reveals that different attenuation laws have to be considered. The \citet{Calzetti00} law is valid for most of the UDF galaxies. Shallower power-law attenuations \citep{Charlot00,LoFaro17} are preferred for most of the GS galaxies. A requirement for flatter attenuation laws is found for galaxies with a very compact dust distribution detected by ALMA, much less extended than the stellar emission observed by HST. The stellar mass estimates are dependent on the choice of the attenuation law, with differences reaching a factor of approximately three when either the stellar continuum or the full SED is used for the fit, assuming different attenuation laws.
The SFRs are robustly estimated as long as the dust emission is considered. We find that the SFR from the fit of the full SED are consistent with the sum of the SFR from the IR and UV.

\begin{acknowledgements}
VB thanks David Elbaz for very fruitful discussions which helped to improve the paper and Tao Wang for providing access to his Good South $\sl Herschel$ data before their publication.
MB acknowledges funding from the FONDECYT regular project 1170618. KM has been supported by the National Science Centre (UMO-2018/30/E/ST9/00082). The project has received funding from Excellence Initiative of Aix-Marseille University - AMIDEX, a French `Investissements d'Avenir' programme.
\end{acknowledgements}

\nocite{*}
\bibliographystyle{aa}
\bibliography{aa-3}
\begin{appendix}
\section{Best fits}
\begin{figure*}
\caption{Best fits obtained for each galaxy of our sample. The observed fluxes (y-axis) are plotted against the effective  wavelength of each filter (x-axis).The solid red line represents  the spectrum of the best fit obtained with  the full SED as described in section 4.3, with the best attenuation law listed in Table \ref{attlaws}, column 4. The dashed blue line represents the spectrum corresponding to  the  best fit of the stellar continuum (section 4.1) extending to IR and submm to show the dust emission predicted. The AGN component, if any, is plotted with a thin green line. The intrinsic, unabsorbed, stellar continuum of each best fit are plotted with a thin  blue line, the  shaded areas indicate the corresponding amount of attenuation (in red for the fit of the full SED, in blue for the fit of the stellar continuum). The data points are plotted with empty blue squares and their 3$\sigma$ error with blue vertical lines.}
\includegraphics[width=0.7\columnwidth] {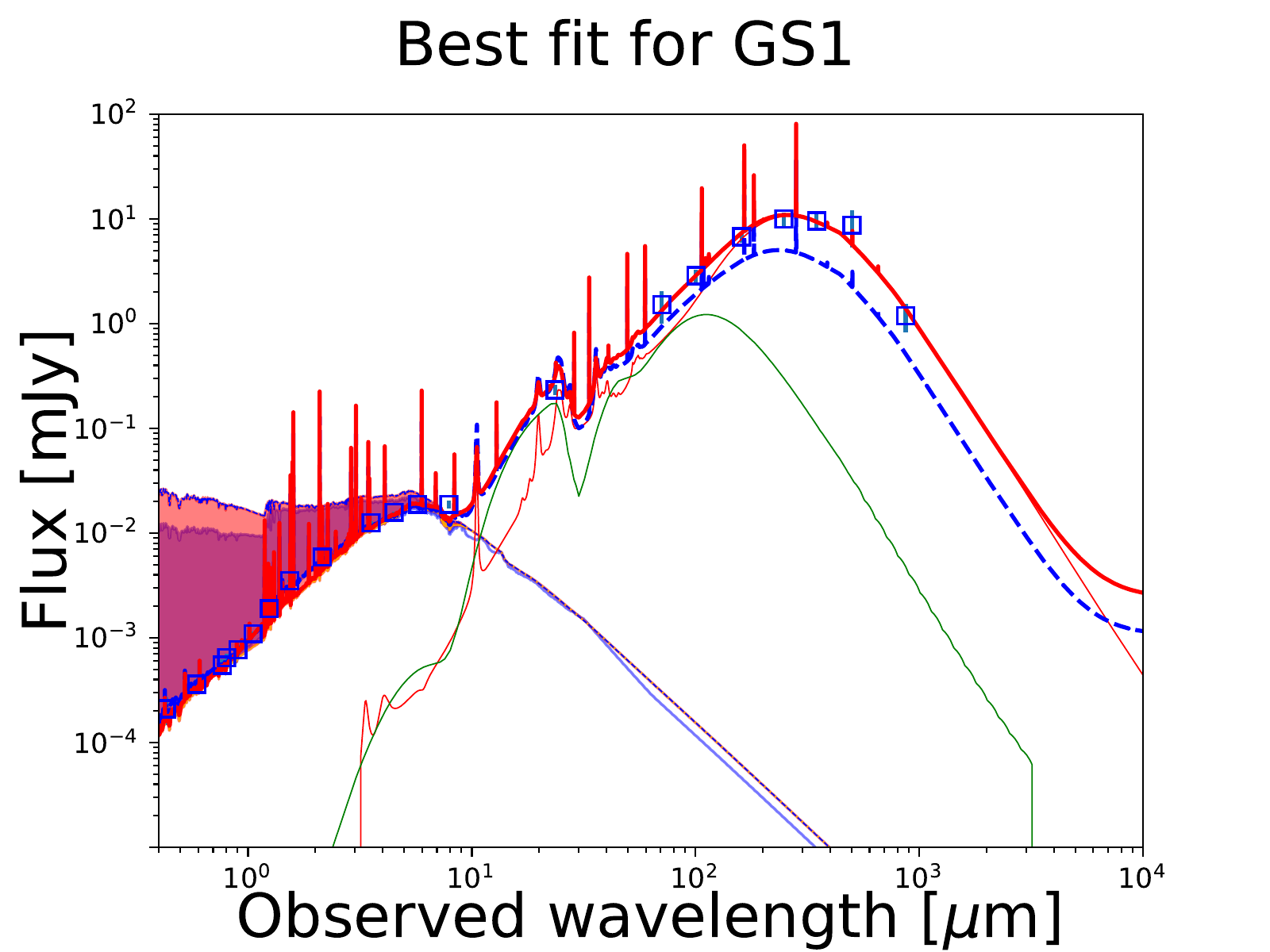}
\includegraphics[width=0.7\columnwidth] {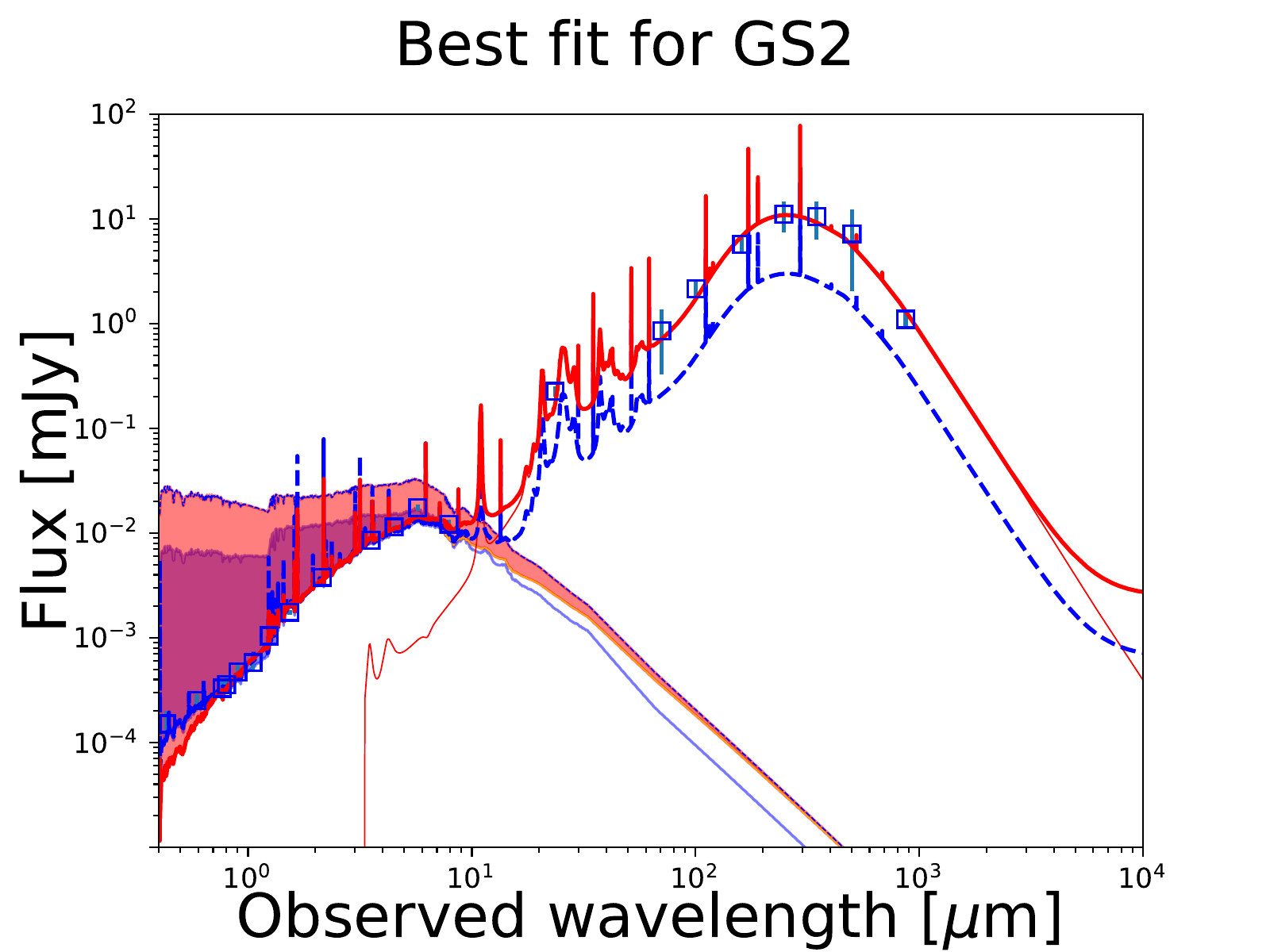}
\includegraphics[width=0.7\columnwidth] {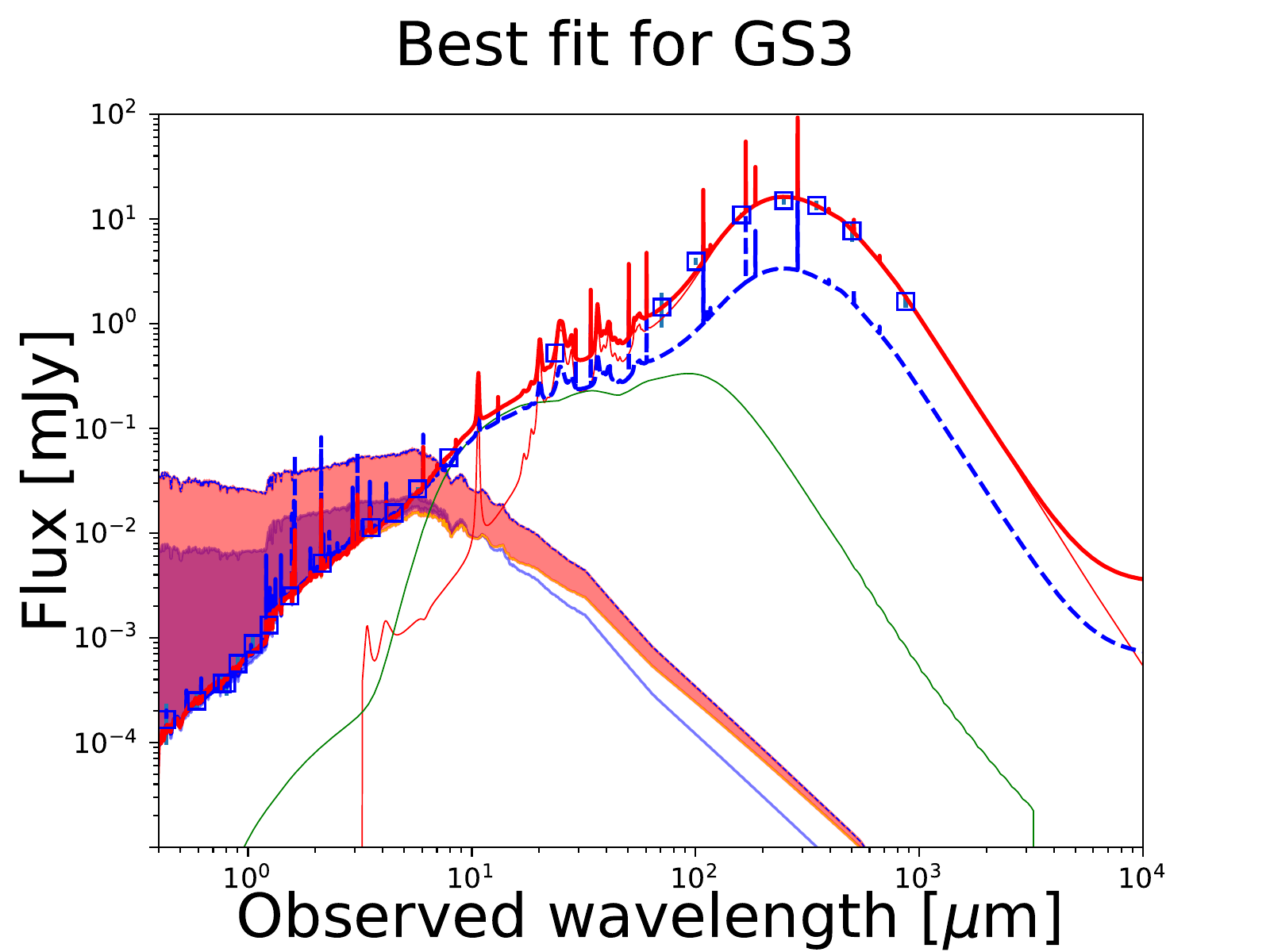}
\includegraphics[width=0.7\columnwidth] {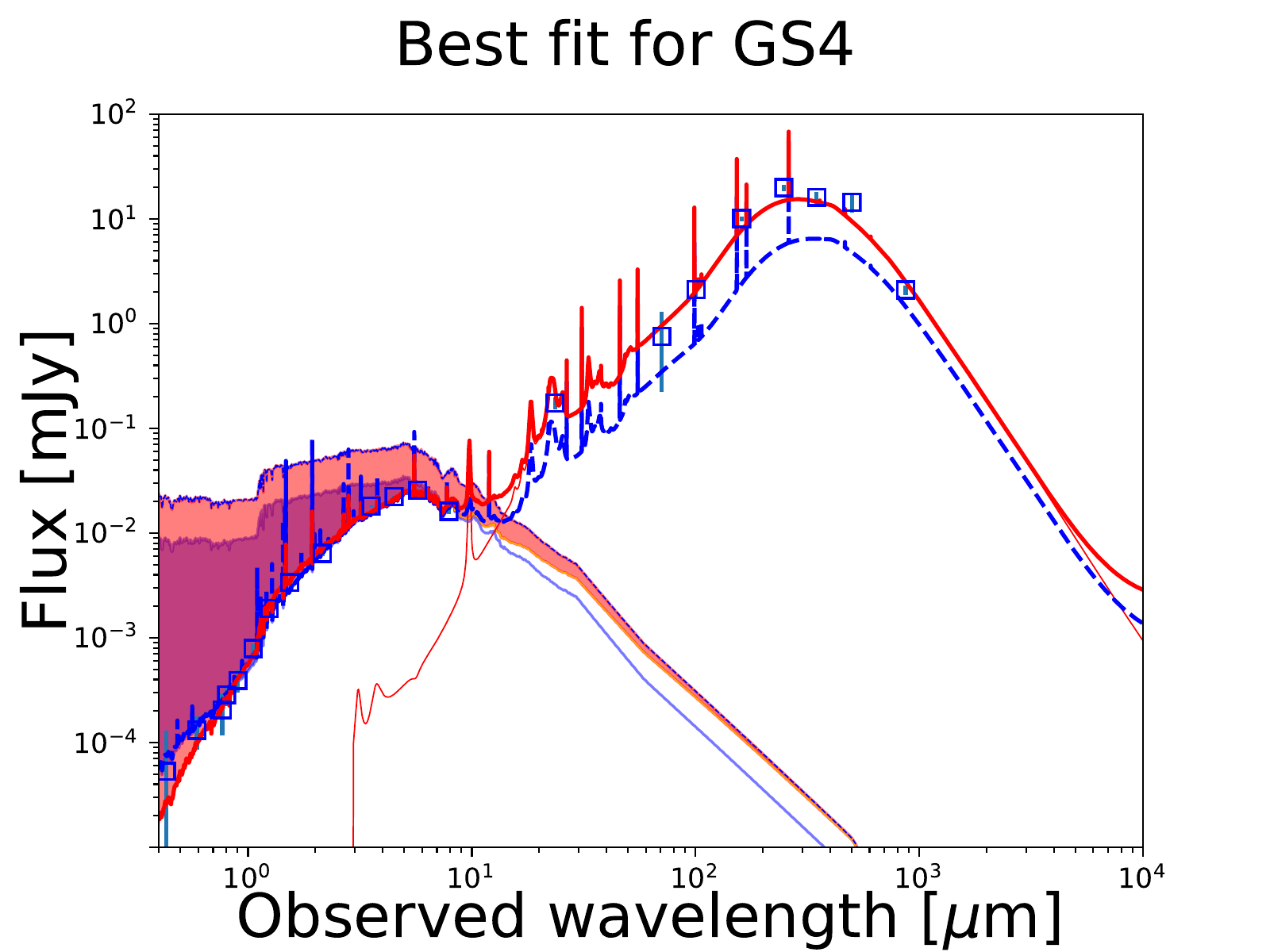}
\includegraphics[width=0.7\columnwidth] {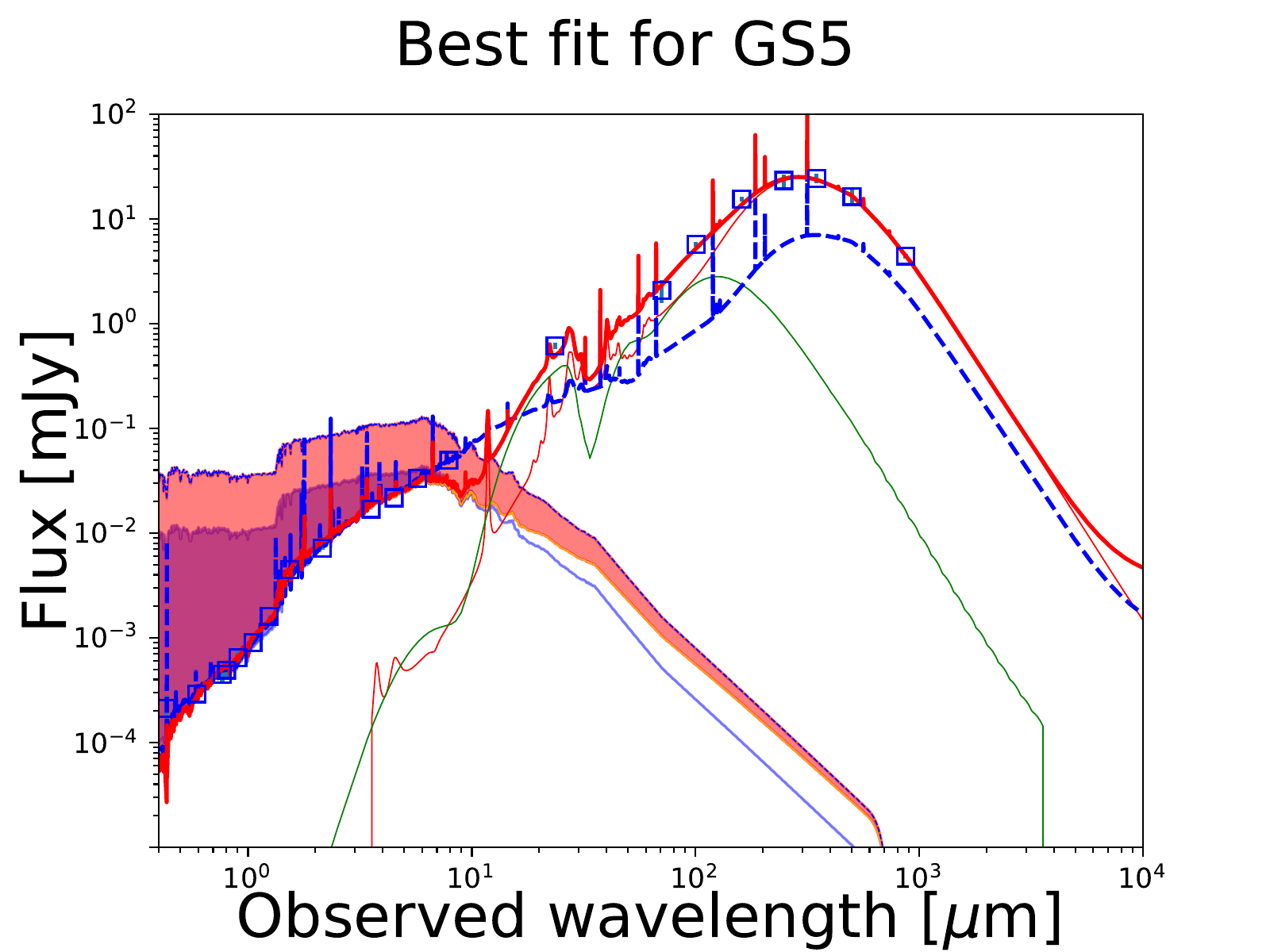}
\includegraphics[width=0.7\columnwidth] {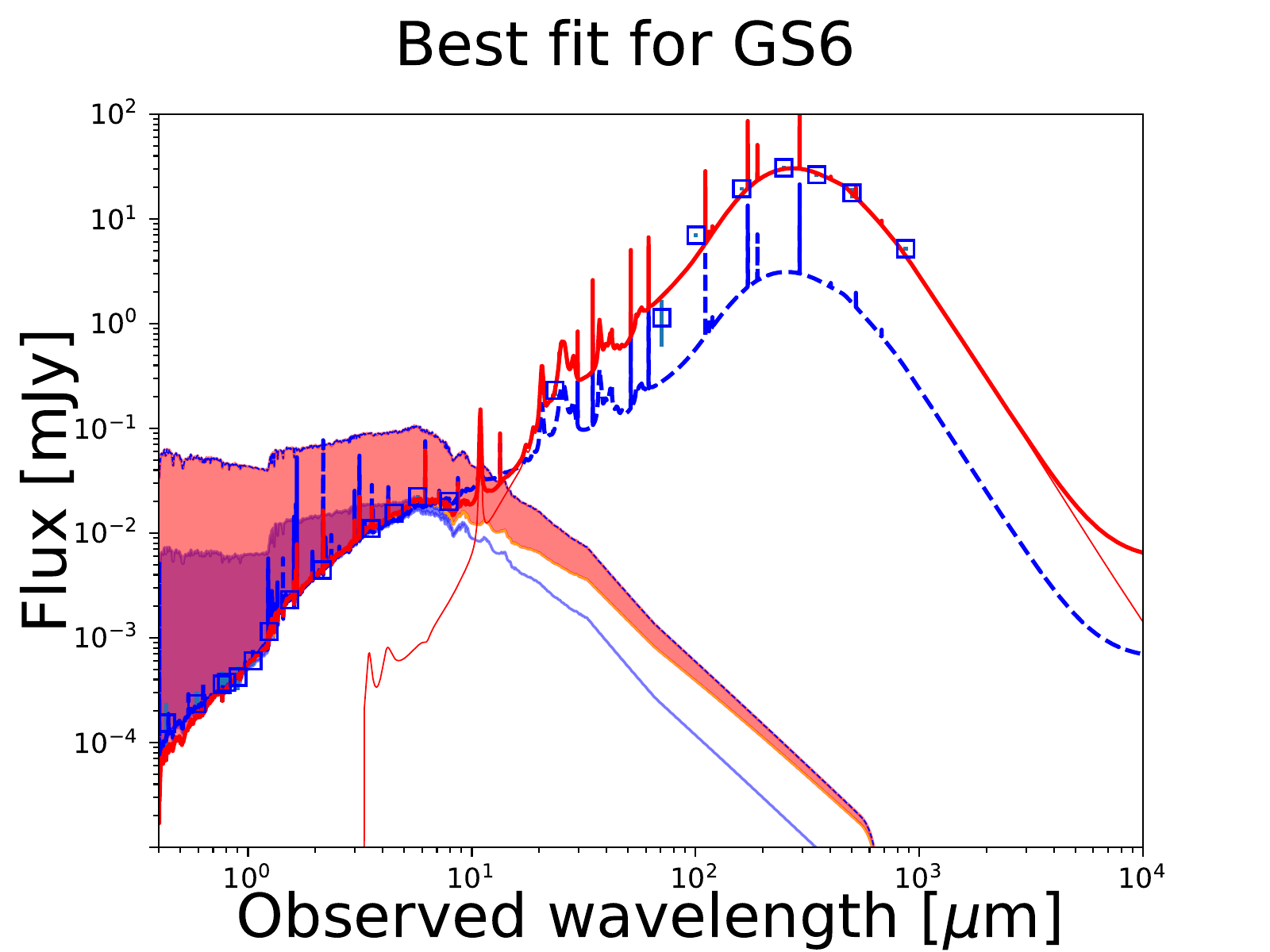}
\includegraphics[width=0.7\columnwidth] {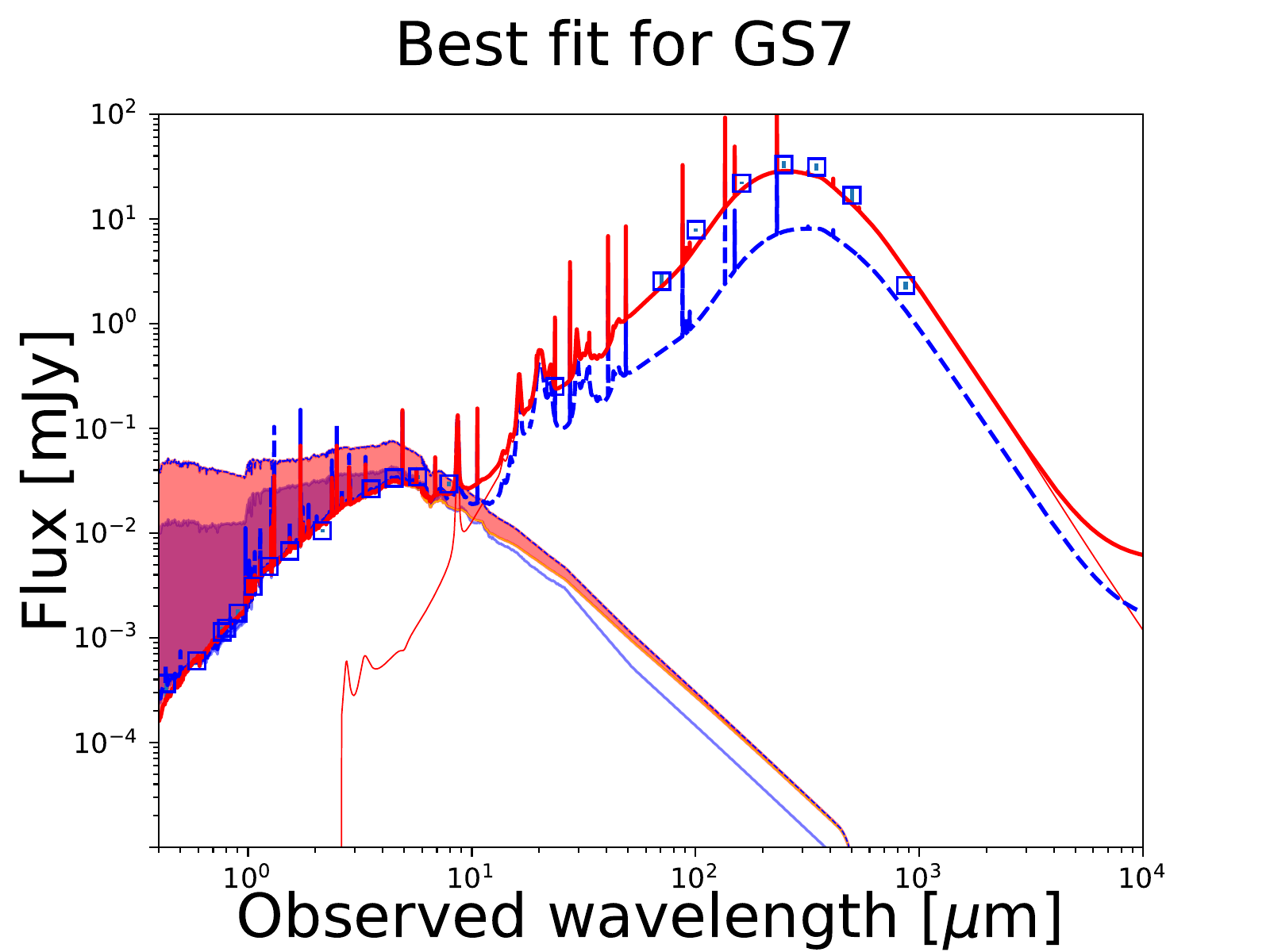}
\includegraphics[width=0.7\columnwidth] {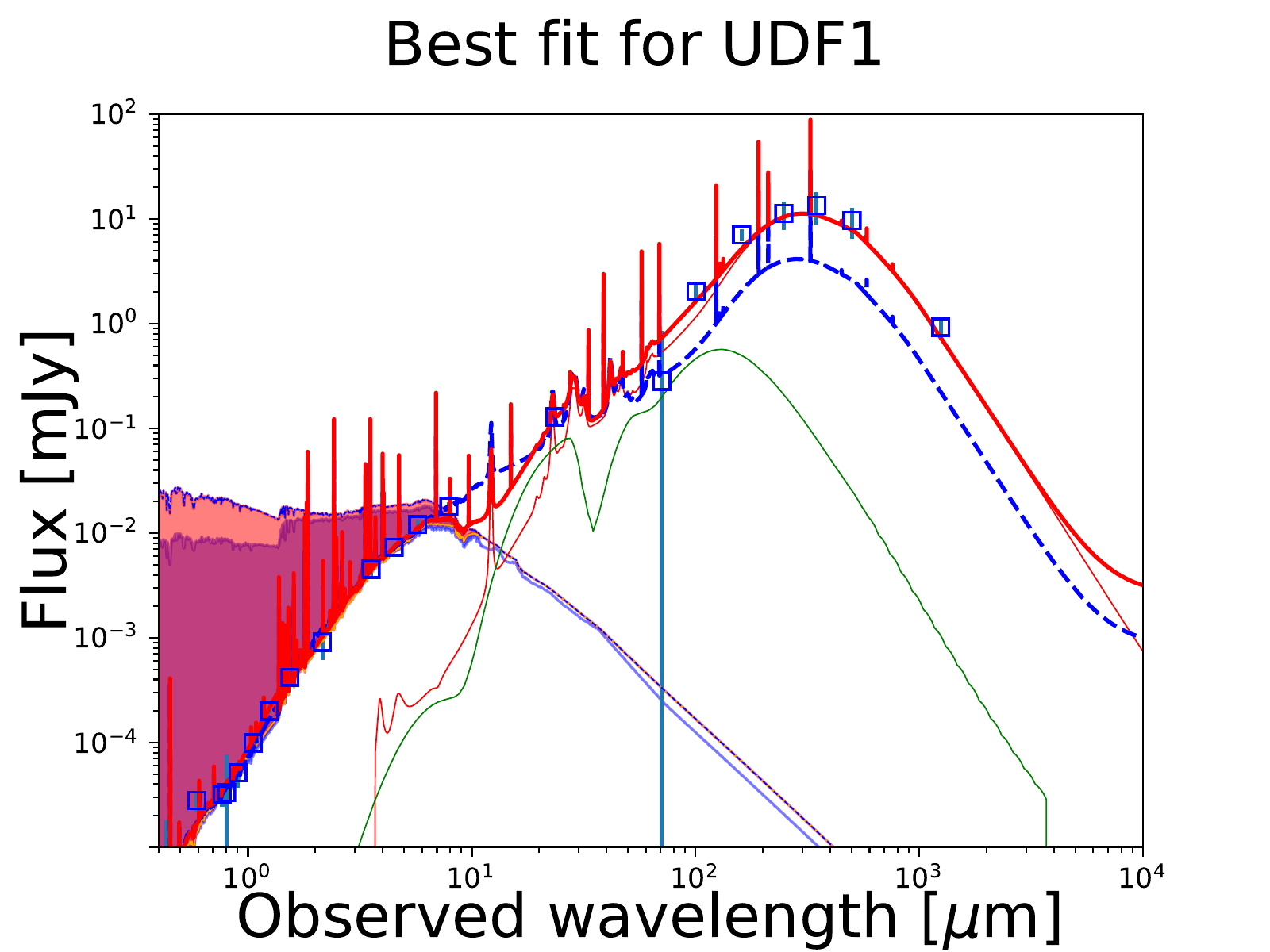}
\includegraphics[width=0.7\columnwidth] {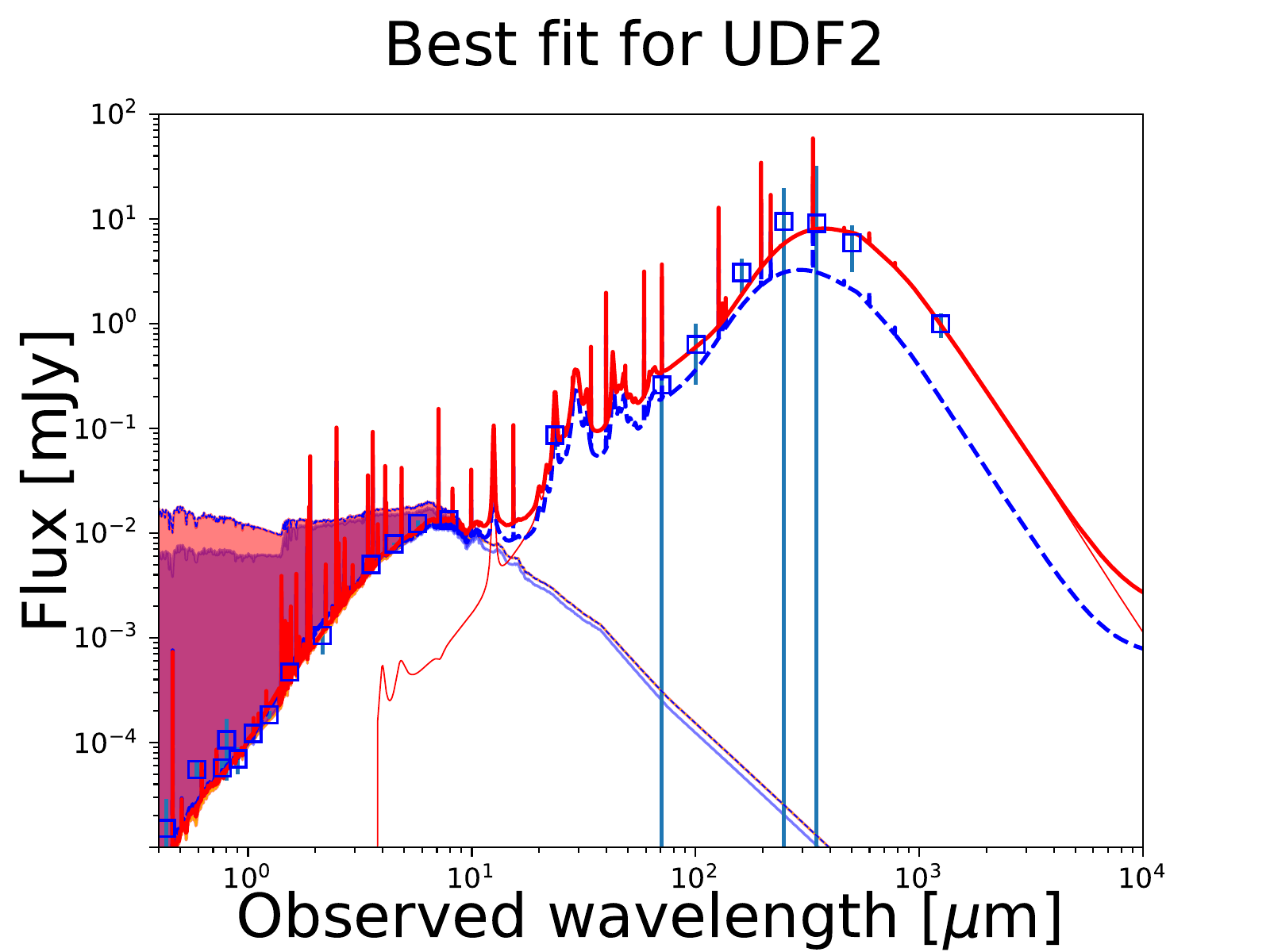}
\includegraphics[width=0.7\columnwidth] {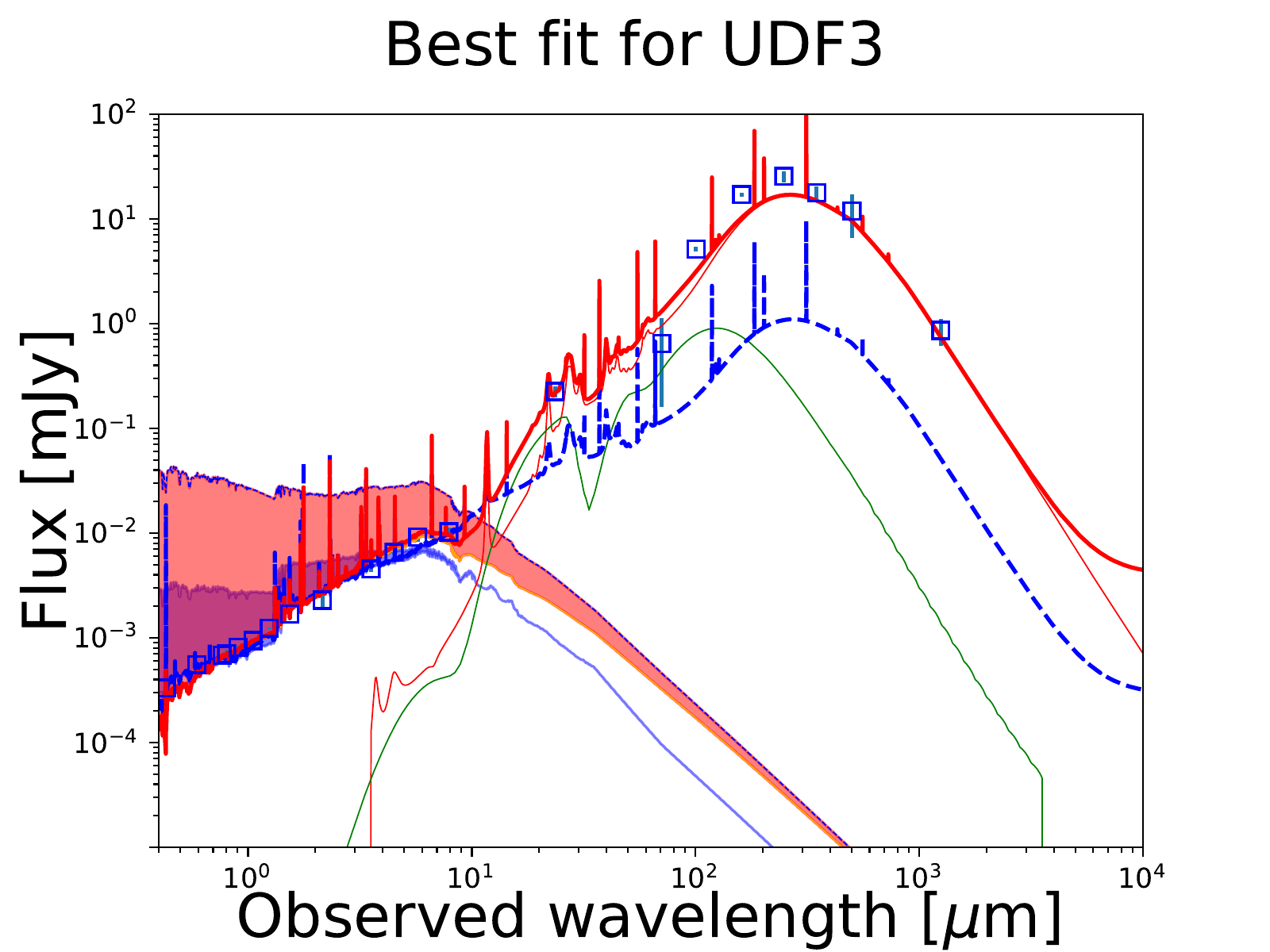}
\includegraphics[width=0.7\columnwidth] {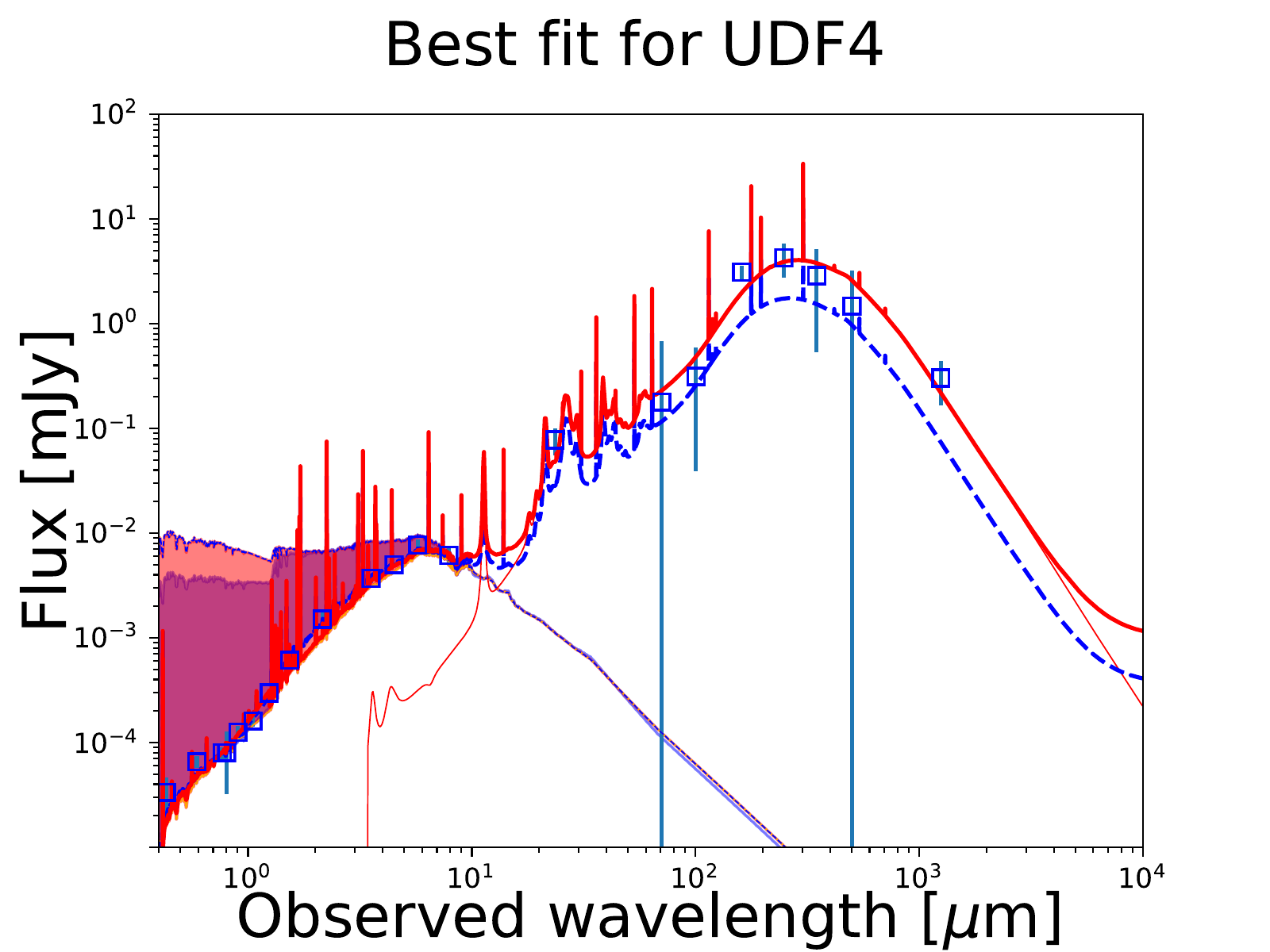}
\includegraphics[width=0.7\columnwidth] {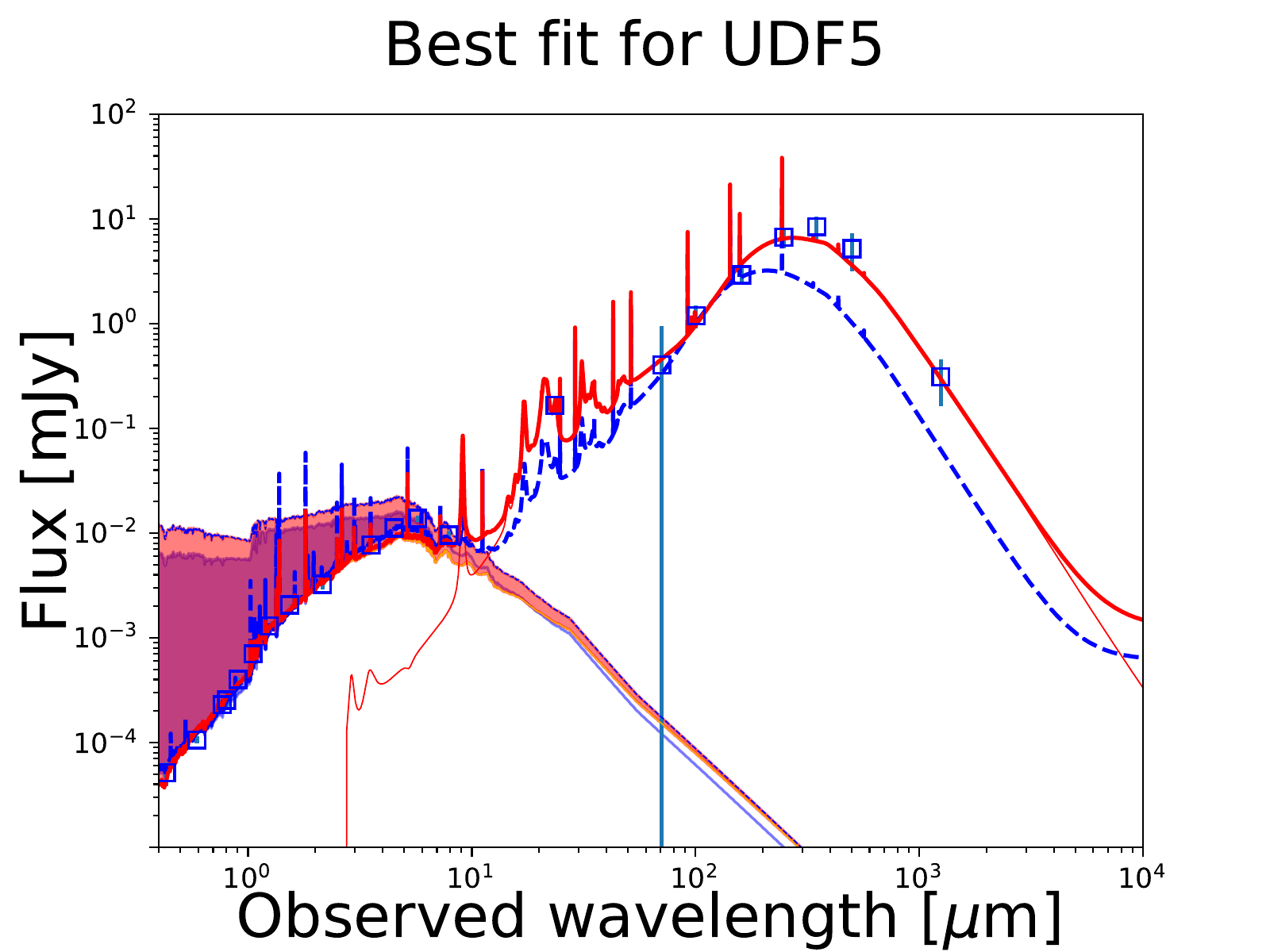}
\includegraphics[width=0.7\columnwidth] {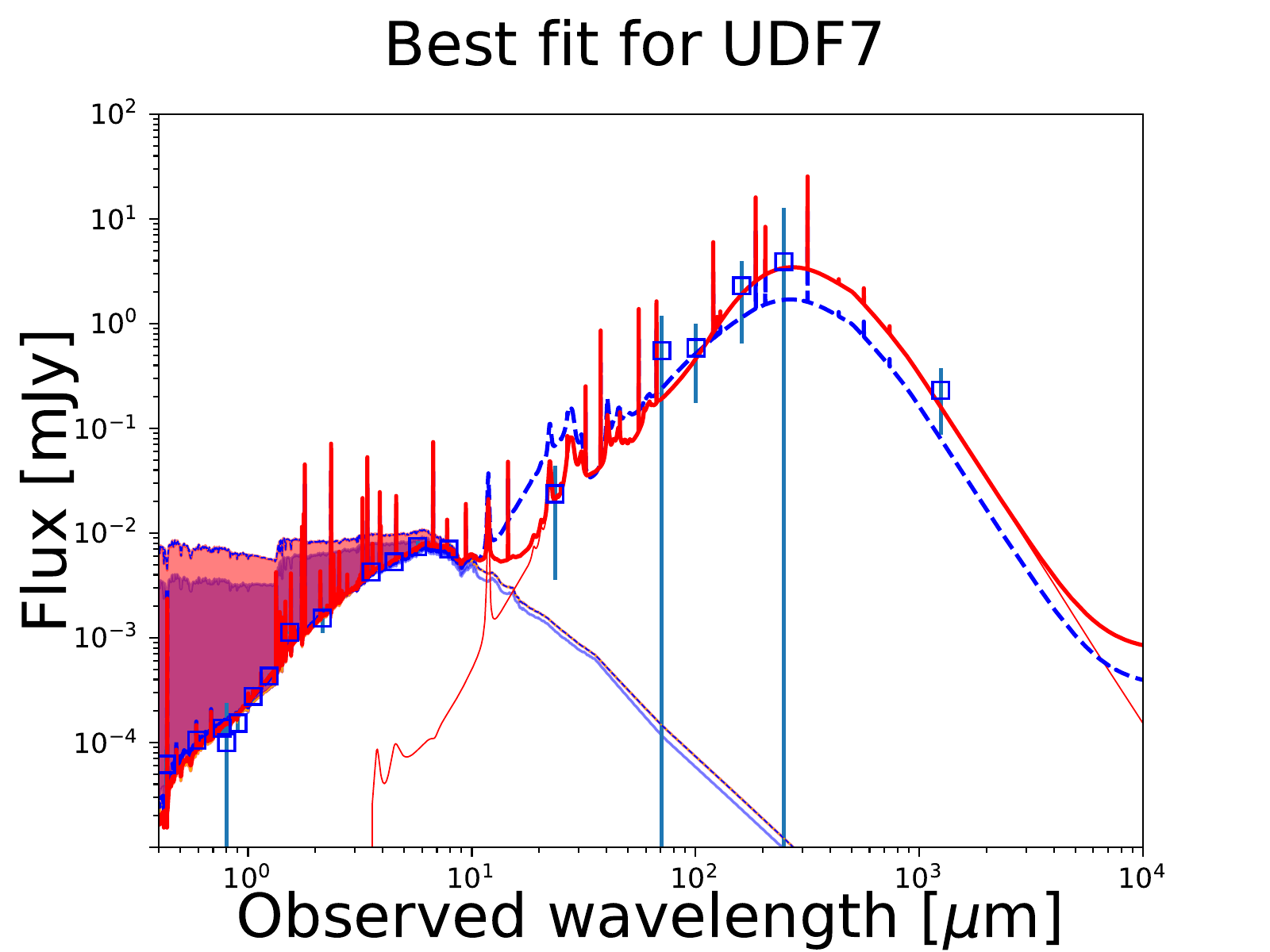}
\includegraphics[width=0.7\columnwidth] {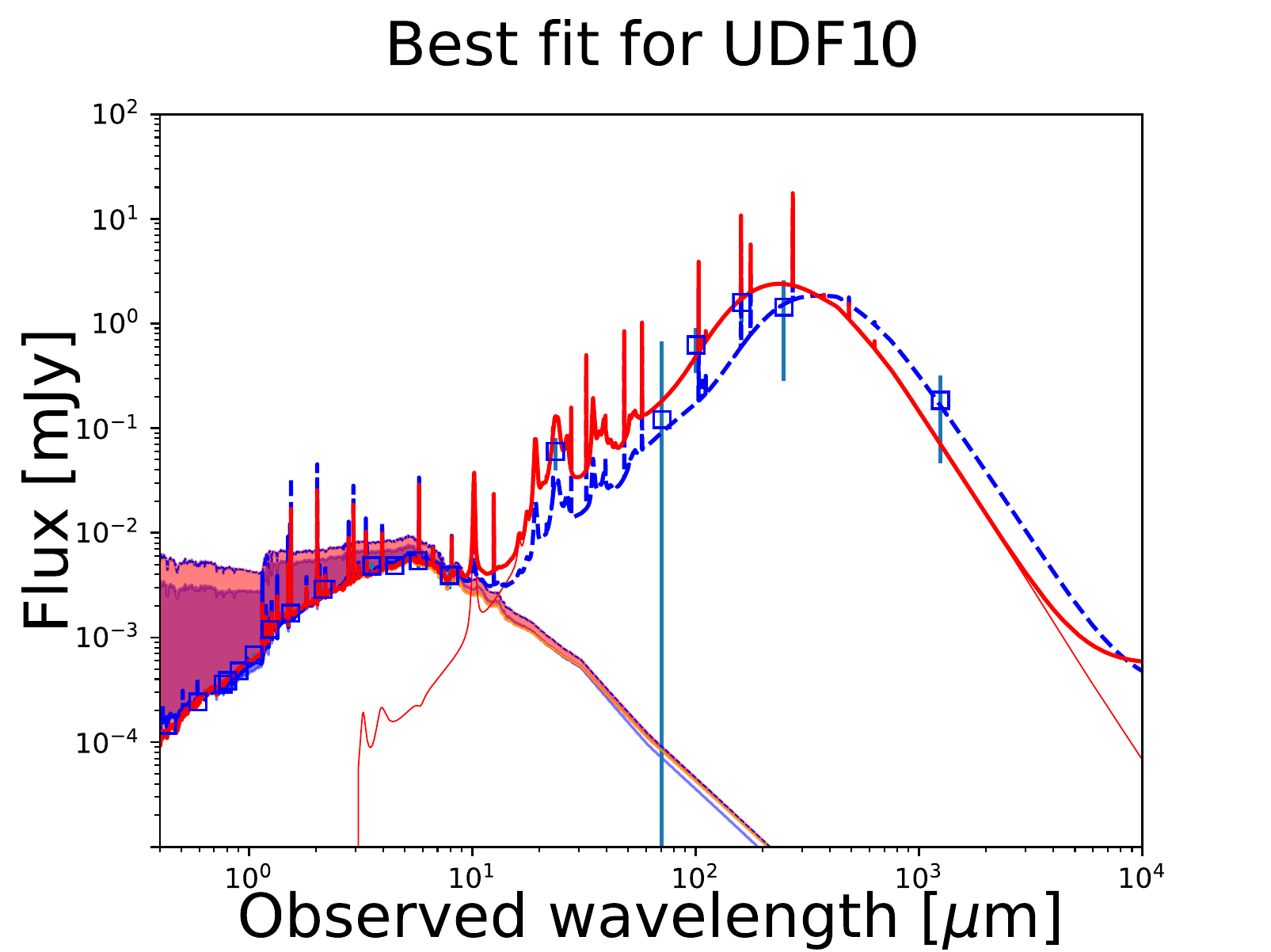}
\includegraphics[width=0.7\columnwidth] {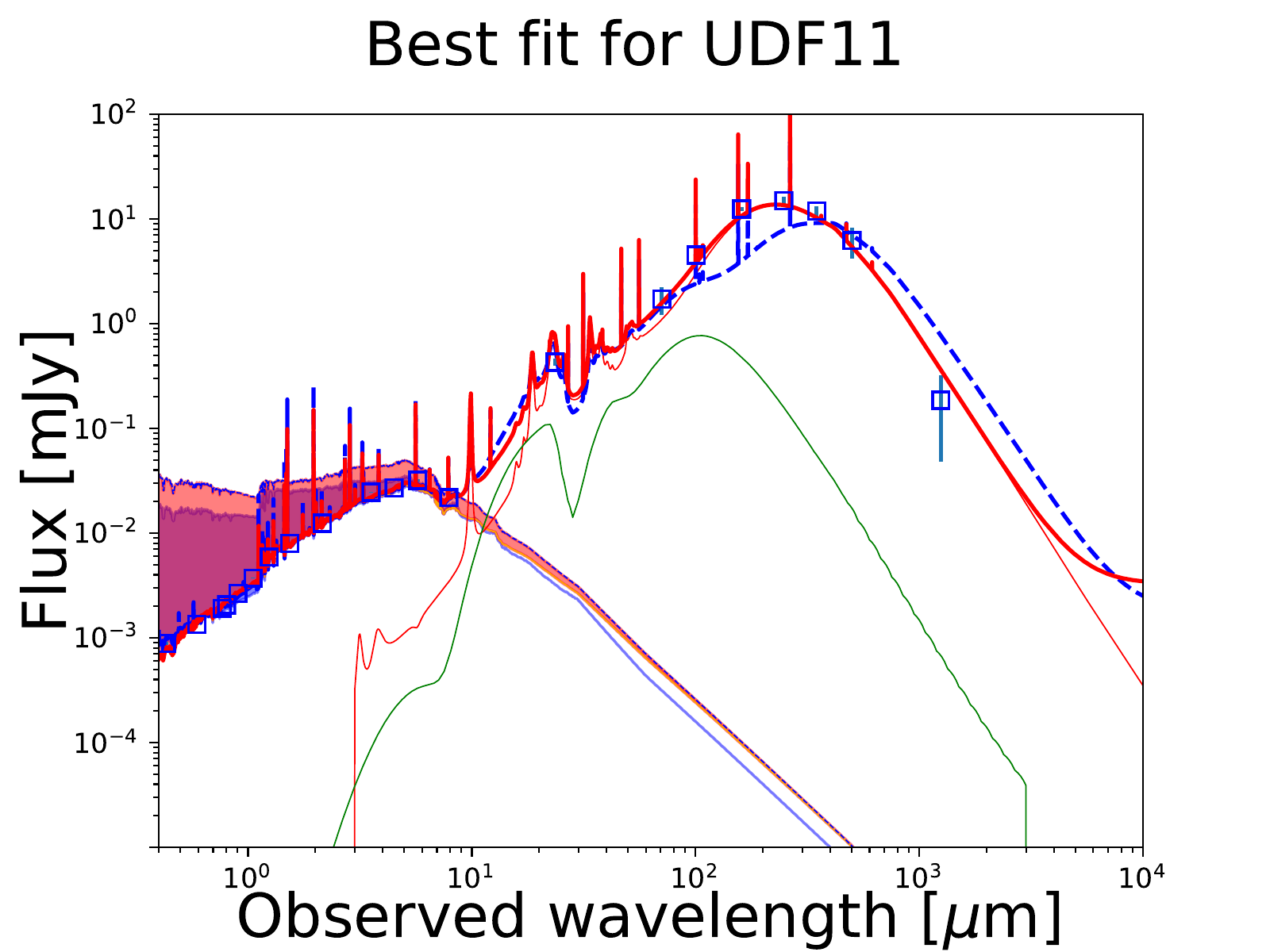}
\end{figure*}
\newpage
\begin{figure*}
\includegraphics[width=0.7\columnwidth] {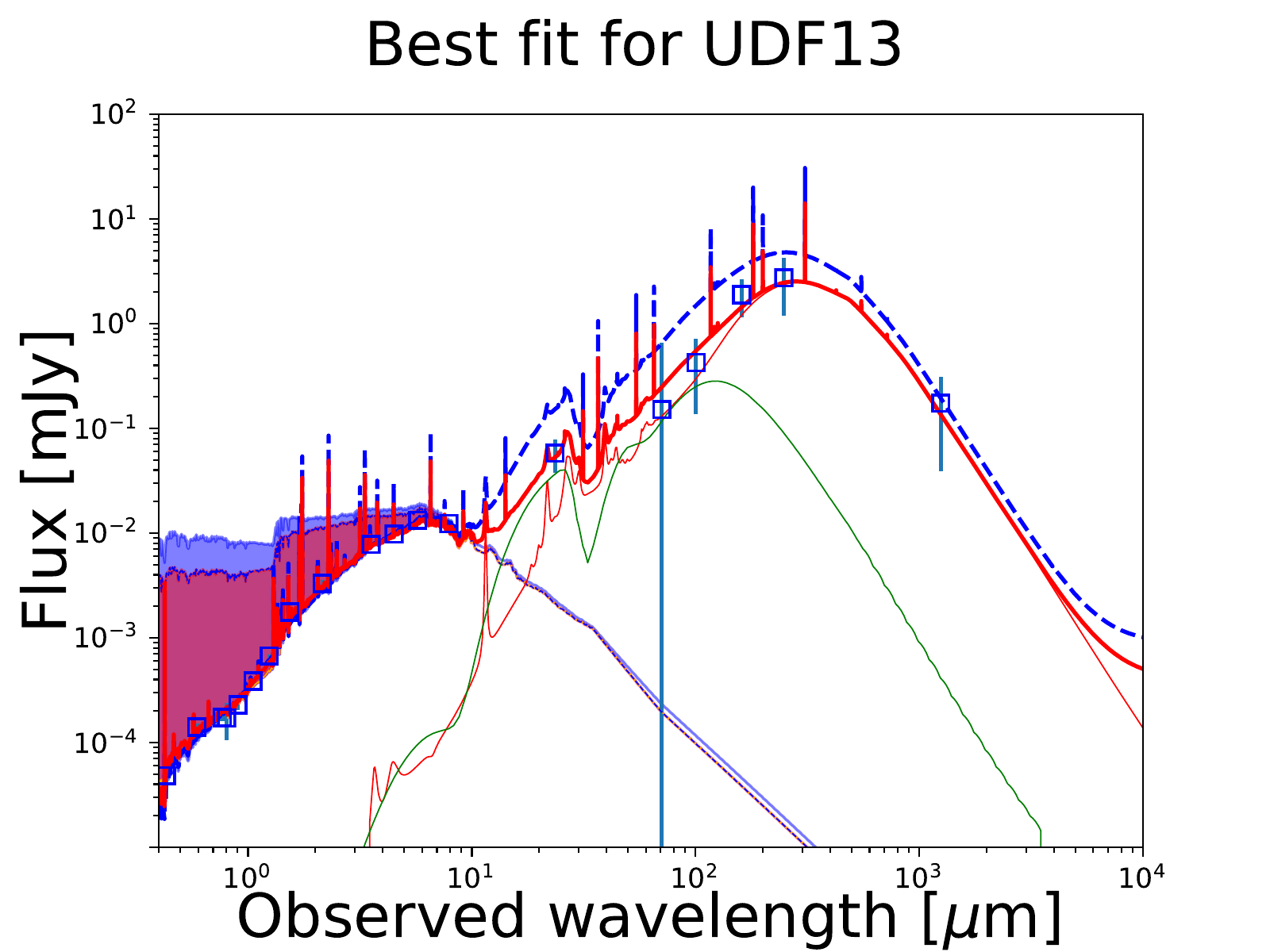}
\includegraphics[width=0.7\columnwidth] {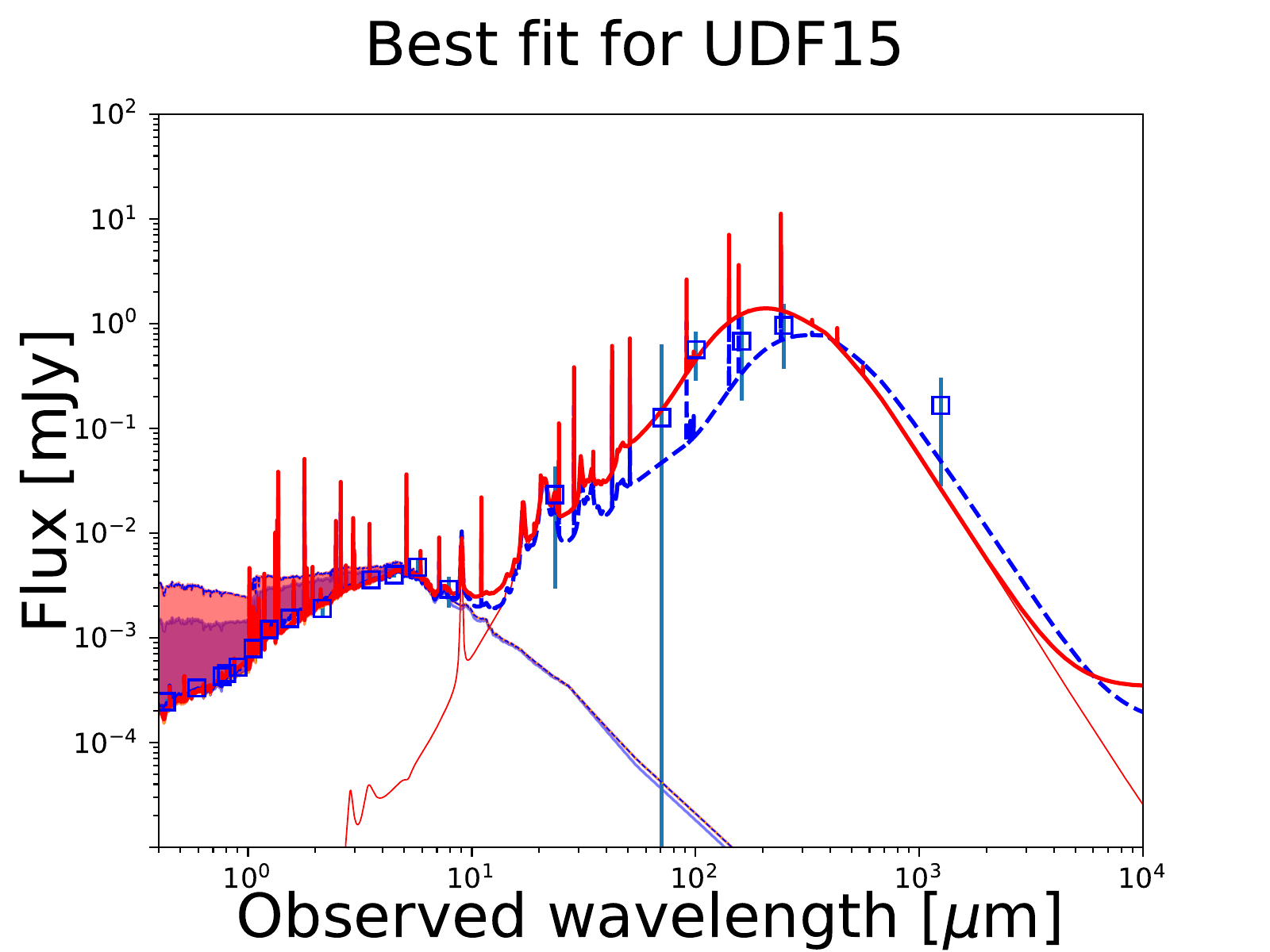}
\label{bestfits}
\label{bestfit}
\end{figure*} 
\end{appendix}
\end{document}